\documentclass[11pt]{article}
\pdfoutput=1
\usepackage{jheppub}

\allowdisplaybreaks[2]

\def\cW{\mathcal{W}}

\def\mint{\int_{-\infty}^\infty\!\cdots\!\int_{-\infty}^\infty}

\newcommand{\be}{\begin{equation}}
\newcommand{\ee}{\end{equation}}
\newcommand{\ba}{\begin{aligned}}
\newcommand{\ea}{\end{aligned}}

\def\bra#1{\left\langle #1 \right|}
\def\ket#1{\left| #1 \right\rangle}

\def\({\left(}
\def\){\right)}

\DeclareMathOperator{\Tr}{Tr}

\DeclareMathOperator{\Det}{Det}

\DeclareMathOperator{\Pexp}{Pexp}

\newcommand{\ri}{{\rm i}}
\newcommand{\rd}{{\rm d}}

\def\cob{\delta}

\newcommand{\hf}{\frac{1}{2}}

\def\til#1{\widetilde{#1}}

\def\del{\partial}

\def\bra{\langle}
\def\ket{\rangle}

\def\la{\lambda}

\def\ka{\kappa}
\def\h#1{\widehat{#1}}

\def\ga{\gamma}
\def\Ga{\Gamma}

\newcommand{\Om}{\Omega}

\def\rt#1{\sqrt{#1}}
\def\sitarel#1#2{\mathrel{\mathop{\kern0pt #1}\limits_{#2}}}

\newdimen\tableauside\tableauside=1.0ex
\newdimen\tableaurule\tableaurule=0.4pt
\newdimen\tableaustep
\def\phantomhrule#1{\hbox{\vbox to0pt{\hrule height\tableaurule width#1\vss}}}
\def\phantomvrule#1{\vbox{\hbox to0pt{\vrule width\tableaurule height#1\hss}}}
\def\sqr{\vbox{%
  \phantomhrule\tableaustep
  \hbox{\phantomvrule\tableaustep\kern\tableaustep\phantomvrule\tableaustep}%
  \hbox{\vbox{\phantomhrule\tableauside}\kern-\tableaurule}}}
\def\squares#1{\hbox{\count0=#1\noindent\loop\sqr
  \advance\count0 by-1 \ifnum\count0>0\repeat}}
\def\tableau#1{\vcenter{\offinterlineskip
  \tableaustep=\tableauside\advance\tableaustep by-\tableaurule
  \kern\normallineskip\hbox
    {\kern\normallineskip\vbox
      {\gettableau#1 0 }%
     \kern\normallineskip\kern\tableaurule}%
  \kern\normallineskip\kern\tableaurule}}
\def\gettableau#1{\ifnum#1=0\let\next=\null\else
\squares{#1}\let\next=\gettableau\fi\next}

\title{Instanton Corrections of
1/6 BPS Wilson Loops in ABJM Theory}

\author{Kazumi Okuyama}

\affiliation{Department of Physics, Shinshu University, Matsumoto 390-8621, Japan}

\emailAdd{kazumi@azusa.shinshu-u.ac.jp}

\abstract{We study instanton corrections to the vacuum expectation value (VEV) of
1/6 BPS Wilson loops in ABJM theory from the Fermi gas approach.
We mainly consider Wilson loops in the fundamental representation
and winding Wilson loops, but we also initiate the study of Wilson loops
with two boundaries.  
We find that the membrane instanton corrections to the Wilson loop VEV are
determined by the refined topological string in the Nekrasov-Shatashvili limit, 
and the pole cancellation 
mechanism between membrane instantons and worldsheet instantons
works also in the Wilson loop VEVs
as in the case
of the partition functions.
}

\begin{document}

\maketitle

\renewcommand{\thefootnote}{\arabic{footnote}}
\setcounter{footnote}{0}
\setcounter{section}{0}

\section{Introduction}
The AdS/CFT correspondence \cite{Maldacena:1997re} has been tested in
many examples and we believe that holography
is one of the fundamental principles of quantum gravity
and string theory.
The localization computation in supersymmetric field theories \cite{Pestun:2007rz}
opened a door to a new era and
we are now in a stage to study the holographic duality at a precise quantitative level.
In particular, in the past few years we have witnessed a tremendous progress in understanding
the holographic duality between M-theory on $AdS_4\times S^7/\mathbb{Z}_k$
and 3d $\mathcal{N}=6$ $U(N)_k\times U(N)_{-k}$
Chern-Simons-matter theory, known as the ABJM theory \cite{ABJM}.
From the analysis of the partition function of ABJM theory on a three-sphere
obtained by the localization technique \cite{Kapustin:2009kz},
the $N^{3/2}$ scaling law of the degrees of freedom on $N$ M2-branes predicted from 
the gravity side \cite{Klebanov:1996un}
is correctly reproduced from the first principles computation on the field theory side \cite{Drukker:2010nc}.

Moreover, there are exponentially small corrections
to the free energy \cite{Drukker:2010nc,Drukker:2011zy},
corresponding to the M2-brane instantons wrapping some three-cycles inside
$S^7/\mathbb{Z}_k$ in the bulk M-theory side.
The Fermi gas formalism  \cite{MP} is a powerful
technique to study such instanton corrections. 
In this approach, we consider the grand partition function of
ABJM theory by introducing the chemical potential $\mu$ and summing over $N$.
It turns out that the grand partition function is written as a Fredholm determinant
which in turn is interpreted as a system of free fermions on a real line.
The grand potential of ABJM Fermi gas system receives two types of
instanton corrections: worldsheet instanton corrections of order $\mathcal{O}(e^{-4\mu/k})$
and membrane instanton corrections of order $\mathcal{O}(e^{-2\mu})$.
There are also bound states of worldsheet instantons and membrane instantons,
but they can be removed by introducing the ``effective'' chemical potential \cite{HMO3}.
Building on a series of works \cite{HMO1,HMO2,Calvo:2012du,HMO3}, it is finally realized that
the grand potential of ABJM theory is completely determined by
the refined topological string on local $\mathbb{P}^1\times \mathbb{P}^1$ \cite{HMMO}.
In particular, the membrane instanton corrections
are given by
the refined topological free energy in the Nekrasov-Shatashvili limit (NS free energy),
while the worldsheet instanton corrections correspond to
the standard, un-refined topological string.
Although the membrane instantons and the worldsheet instantons separately have poles at
rational values of $k$, those poles are actually canceled by adding the two contributions.
This {\it pole cancellation mechanism} guarantees
that the grand partition function is well-defined for all $k$.
For the special values $k=1,2,4,8$,
the grand partition function can be written
in closed forms in terms of the Jacobi theta functions
\cite{Codesido:2014oua,Grassi:2014uua,Okuyama:2016xke}.

One can also  compute exactly
the vacuum expectation value (VEV) of 1/6 BPS Wilson loops in ABJM theory using
the localization technique \cite{Kapustin:2009kz}.
The study of such BPS Wilson loops
from the Fermi gas approach was initiated in \cite{KMSS} and further developed in \cite{HMMO-Wilson}
particularly for
the 1/2 BPS Wilson loops.
It is found that 1/2 BPS Wilson loops in arbitrary representations
are given by a determinant of hook representations \cite{HMMO-Wilson,Matsuno:2016jjp,HO-1/2}, 
and 1/2 BPS Wilson loops are closely related to the open topological string on 
local $\mathbb{P}^1\times \mathbb{P}^1$ \cite{Marino:2009jd,Grassi:2013qva,HO-1/2,Marino:2016rsq}.
Interestingly, it is observed that there are no ``pure'' membrane instanton corrections in 1/2 BPS Wilson loops
\cite{HMMO-Wilson} 
which turns out to be a consequence of the 
{\it open-closed duality} between the 1/2 BPS Wilson loops and
the grand partition functions of ABJ theory \cite{HO-1/2}.
In particular, there is no pole cancellation between worldsheet instantons and 
membrane instantons in the VEV of  1/2 BPS Wilson loops. 

On the other hand, 1/6 BPS Wilson loops in ABJM theory have no 
direct connection to open topological strings and the 
instanton corrections of 1/6 BPS Wilson loops have not been fully explored in the literature.
In this paper, we initiate a study of the instanton corrections of 1/6 BPS Wilson loops
mainly using the numerical analysis.
We will show that the VEV of 1/6 BPS Wilson loops
can be computed numerically with high precision and
we will study the instanton corrections of 1/6 BPS Wilson loops
from numerical fitting.
We first consider the 1/6 BPS Wilson loops in the fundamental representation.
We find that the membrane instanton corrections
are given by the NS free energy, and 
the pole cancellation mechanism works also for the 1/6 BPS Wilson loops.
We find that, up to regular terms, the 1/6 BPS Wilson loops in the fundamental representation 
is essentially determined by the quantum
volume appearing in the exact quantization condition of the spectrum \cite{Kallen:2013qla,Kallen:2014lsa,Wang:2014ega,Grassi:2014uua,Wang:2015wdy,Hatsuda:2015fxa}.

We also study the 1/6 BPS Wilson loops
with winding number $n\geq2$ and
the 1/6 BPS Wilson loops with two boundaries. 
Our numerical study shows that the perturbative part of
the 1/6 BPS Wilson loops
with winding number $n\geq2$ is different from
the expression obtained in \cite{KMSS}.
We also study the instanton corrections to those Wilson loops
and find evidence that the membrane instanton corrections are
again related to the NS free energy.

This paper is organized as follows.
In section \ref{sec:rev}, we first review
the Fermi gas approach to the 1/6 BPS Wilson loops, and
explain our algorithm to compute them numerically.
In section \ref{sec:pert}, we consider the 
perturbative part (or zero-instanton part) of
the grand canonical VEV of winding Wilson loops.
Our result \eqref{eq:Wn-pert} is different from the one found in \cite{KMSS}
for winding number $n\geq2$. In section \ref{sec:fund},
we study the instanton corrections to the fundamental Wilson loop numerically.
We find that the grand canonical VEV of fundamental Wilson loop 
is closely related to the quantum volume \eqref{eq:W1-final}. 
In section \ref{sec:WKB}, we study the WKB expansion of
the fundamental Wilson loop. Using the Pad\'e approximation
we confirm 
that the membrane instanton corrections are correctly reproduced
from the WKB expansion. 
In section \ref{sec:wind}, we consider winding Wilson loops with winding number $n\geq2$. 
We find  that the 
membrane instanton corrections for the winding number $n=2$
are also related to the NS free energy.
We also find the first few worldsheet instanton corrections for
the winding numbers $n=3,4$. In section \ref{sec:two-b}, we study the Wilson loops with
two boundaries. We show that 
the VEV of Wilson loops with
two boundaries can be systematically computed by 
constructing a sequence of functions, in a similar manner as
the winding Wilson loops.
Then we study the instanton corrections of Wilson loops with
two boundaries. Again, the membrane instanton corrections to the imaginary 
part of Wilson loop is related to the NS free energy.
Finally, we conclude in section \ref{sec:conclude}.
In Appendix \ref{app:inst}, we summarize the instanton corrections for some integer
values of $k$.
\section{$1/6$ BPS Wilson loops from Fermi gas approach}\label{sec:rev}
In this paper, we will consider the VEV
 of 
1/6 BPS Wilson loops in ABJM theory on $S^3$, first constructed in \cite{DPY,CW,RSY}
\begin{align}
 W_R^{(1/6)}=\Tr_R \Pexp \left[ \int \! \rd s \( \ri A_\mu \dot{x}^\mu+\frac{2\pi}{k} |\dot{x}| M^I_J C_I \bar{C}^J \) \right].
\label{eq:def-W}
\end{align}
Here $x^\mu(s)$ parametrizes a great circle of $S^3$,  $C_I$ ($I=1,2,3,4$) are the scalar fields in the bi-fundamental chiral multiplets, and  
$M^I_J$ is a constant matrix which can be brought to the form $M=\text{diag}(1,1,-1,-1)$
in a suitable choice of basis. 
$A_\mu$ 
is the gauge field of one of the $U(N)$ factor of the gauge group $U(N)_k\times U(N)_{-k}$ of ABJM theory.
On the other hand, the construction of 1/2 BPS Wilson loops 
involves the gauge fields in both of the $U(N)$ factors and 
utilizes the supergroup
structure $U(N|N)$ \cite{DT}.

The VEV of such 1/6 BPS Wilson loops can be reduced to
a matrix model by the supersymmetric localization \cite{Kapustin:2009kz}
\begin{align}
\begin{aligned}
 \bra W_R^{(1/6)}\ket_N&=\frac{1}{(N!)^2}
\int\prod_{i=1}^N\frac{d\mu_id\nu_i}{(2\pi)^2}
e^{\frac{\ri k}{4\pi}(\mu_i^2-\nu_i^2)}\frac{\prod_{i<j}\big(2\sinh\frac{\mu_i-\mu_j}{2}\big)^2\big(2\sinh\frac{\nu_i-\nu_j}{2}\big)^2}{\prod_{i,j}2\cosh\frac{\mu_i-\nu_j}{2}} \Tr_R U,
\end{aligned}
\label{eq:WR-def}
\end{align}
where $U$ corresponds to the holonomy in one of the gauge group $U(N)$
\begin{align}
 U=\text{diag}(e^{\mu_1},\cdots, e^{\mu_N}).
\label{eq:U-hol}
\end{align}
One can also consider the Wilson loop in the other $U(N)$ factor,
but it is related to \eqref{eq:WR-def} by the complex conjugation (or $k\to-k$).
In this paper we consider the {\it un-normalized VEV}, i.e. we do not divide
\eqref{eq:WR-def} by the partition function $Z(N,k)$.

As shown in \cite{KMSS},
the VEV of 1/6 BPS Wilson loops in the grand canonical picture can be written as 
a Fermi gas system. 
Let us first recall the Fermi gas description of the partition function of ABJM theory
on $S^3$ \cite{MP}. Introducing the fugacity $\ka=e^\mu$ with chemical potential $\mu$,
we define the grand canonical partition function by summing over $N$
\begin{align}
 \Xi(\ka,k)=1+\sum_{N=1}^\infty \ka^N Z(N,k).
\end{align}
It turns out that
the grand partition function can be written as a Fredholm determinant
\begin{align}
 \Xi(\ka,k)=\Det(1+\ka\rho),
\end{align}
where the density matrix $\rho$ is given by
\begin{align}
 \rho=\frac{1}{2\cosh\frac{x}{2}}\frac{1}{2\cosh\frac{p}{2}}.
\end{align}
Here $x$ and $p$ obeys
the canonical commutation relation
\begin{align}
 [x,p]=i\hbar,
\label{eq:comm}
\end{align}
and the Chern-Simons level $k$ and the Planck constant $\hbar$ are related
by
\begin{align}
 \hbar=2\pi k.
\label{eq:hbar}
\end{align}
Similarly,
we can define the grand canonical VEV of a general operator $\mathcal{O}$ by
\begin{align}
 \bra \mathcal{O}\ket^{\text{GC}}=\frac{1}{\Xi(\ka,k)}\sum_{N=1}^\infty \ka^N \bra \mathcal{O}\ket_N.
\end{align}
The crucial observation in \cite{KMSS,HMMO-Wilson} is that
the grand canonical VEV of the holonomy $U$ in \eqref{eq:U-hol}
corresponds to the insertion of a quantum mechanical operator $W$
\begin{align}
 W=e^{\frac{x+p}{k}}
\end{align}
into the Fredholm determinant.
More generally, the grand canonical VEV of operator $\det f(U)$
for some function $f$ is written as
\begin{align}
 \bra \det f(U) \ket^{\text{GC}}=
\Biggl\bra \prod_{i=1}^N f(e^{\mu_i}) \Biggr\ket^{\text{GC}}
 =\frac{\Det(1+\ka\rho f(W))}{\Det(1+\ka\rho)}.
 \label{eq:det-f}
\end{align}
For instance, setting $f$ to
\begin{align}
 f(e^{\mu_i})=1+\varepsilon e^{n\mu_i}
\end{align}
and picking up the term of order $\mathcal{O}(\varepsilon)$,
we find that the grand canonical VEV of 1/6 BPS Wilson loops with winding number $n$
is given by
\begin{align}
 \bra \Tr U^n\ket^\text{GC}=\Tr\left(RW^n\right),
\label{eq:Un-GC}
\end{align}
where we defined $R$ as
\begin{align}
 R=\frac{\ka\rho}{1+\ka\rho}.
 \label{eq:R-def}
\end{align}
Once we know the grand canonical VEV, one can easily find
the canonical VEV with fixed $N$ by
\begin{align}
 \bra \Tr U^n\ket_N=\int_{-\pi\ri}^{\pi\ri}\frac{d\mu}{2\pi \ri }e^{-N\mu}
\Xi(\mu,k)\bra \Tr U^n\ket^\text{GC}.
\label{eq:2pi-int}
\end{align}
For the special case $n=1$, $\bra\Tr U\ket$ corresponds to the
1/6 BPS Wilson loop in the fundamental representation.

\subsection{Computation of trace}\label{sec:exact}

From the small $\ka$ expansion of the grand canonical VEV \eqref{eq:Un-GC}
\begin{align}
\begin{aligned}
 \bra \Tr U^n\ket^\text{GC}&=\sum_{\ell=1}^\infty (-1)^{\ell-1}\ka^\ell\Tr(\rho^\ell W^n),
\label{eq:exp-trU}
\end{aligned}
\end{align}
one can see that the Wilson loop VEV $\bra \Tr U^n\ket_N$
at fixed $N$ can be computed from the traces $\Tr(\rho^\ell W^n)$ and
$\Tr\rho^\ell$ with
$\ell=1,\cdots, N$. 
Note that the trace $\Tr\rho^\ell$  appears in the computation
of the partition function as well. One can compute $\Tr\rho^\ell$ 
by applying the Tracy-Widom lemma \cite{TW}.
Thus our remaining task is to compute
the trace $\Tr(\rho^\ell W^n)$  with $W^n$ insertion systematically.
As we will show below, one can also apply the Tracy-Widom lemma
to the computation of the trace $\Tr(\rho^\ell W^n)$.

First we observe that, using the Baker-Campbell-Hausdorf formula,
$W^n$ is written as 
\begin{align}
 W^n=e^{\frac{n(x+p)}{k}}=e^{\frac{np}{k}}e^{\frac{n(x+n\pi\ri)}{k}}.
\label{eq:W-BCH}
\end{align}
Noticing that the operator $e^{\frac{np}{k}}$
shifts $x$ by $2\pi\ri n$,
one can show that the matrix element of $\rho W^n$
is given by\footnote{We use the following normalization of quantum mechanical states
\begin{align}
 \bra x|y\ket=\cob(x-y),\quad
\bra p|p'\ket=\cob(p-p'),\quad
\bra x|p\ket=\frac{1}{\rt{2\pi\hbar}}e^{\frac{\ri px}{\hbar}},\quad
\bra x|\frac{1}{2\cosh\frac{p}{2}}|y\ket=\frac{1}{\hbar}\frac{1}{2\cosh\frac{x-y}{2k}}.
\end{align}}
\begin{align}
 \bra x|\rho W^n|y\ket&= \bra x|\rho|y+2\pi \ri n\ket e^{\frac{n(y+n\pi\ri)}{k}}, 
\end{align}
where the matrix element of $\rho$ is given by
\begin{align}
\begin{aligned}
 \bra x|\rho|y\ket&=\frac{1}{\hbar}\frac{1}{2\cosh\frac{x}{2}}\frac{1}{2\cosh\frac{x-y}{2k}}
=\frac{1}{\hbar}\frac{1}{2\cosh\frac{x}{2}}\frac{e^{\frac{x+y}{2k}}}{e^{\frac{x}{k}}+e^{\frac{y}{k}}}.
\end{aligned}
\end{align} 
This is exactly the form of kernel to which we can apply
the Tracy-Widom lemma \cite{TW}. 
As shown in \cite{TW},
the matrix element $\bra x|\rho^\ell|y\ket$ can be written as
\begin{align}
 \bra x|\rho^\ell|y\ket=\frac{1}{\hbar}\frac{1}{2\cosh\frac{x}{2}}
\frac{e^{\frac{x+y}{2k}}}{e^{\frac{x}{k}}+(-1)^{\ell-1}e^{\frac{y}{k}}}
\sum_{j=0}^{\ell-1}(-1)^j\phi_j(x)\phi_{\ell-1-j}(y),
\end{align}
where  $\phi_j(x)$ is determined recursively
\begin{align}
 \phi_{j+1}(x)=e^{-\frac{x}{2k}}\int\frac{dy}{\hbar}
\frac{1}{2\cosh\frac{x-y}{2k}}\frac{e^{\frac{y}{2k}}}{2\cosh\frac{y}{2}}\phi_j(y),\qquad
\phi_0(x)=1.
\label{eq:phij}
\end{align}
Since the trace $\Tr(\rho^\ell W^n)$
with insertion of $W^n$ is given by
the integral of $\bra x|\rho^\ell|y\ket$ 
with the replacement
$y\to x+2\pi\ri n$
\begin{align}
 \Tr(\rho^\ell W^n)=\int dx \bra x|\rho^\ell|x+2\pi \ri n\ket e^{\frac{n(x+n\pi\ri)}{k}},
\end{align}
this trace can also be written in terms of the functions
$\phi_j(x)$ in \eqref{eq:phij}
\begin{align}
 \Tr(\rho^\ell W^n)=\frac{e^{\frac{\pi\ri n(n+1)}{k}}}{1+(-1)^{\ell-1}e^{\frac{2\pi\ri n}{k}}}\int\frac{dx}{\hbar}\frac{e^{\frac{nx}{k}}}{2\cosh\frac{x}{2}}
\sum_{j=0}^{\ell-1}(-1)^j\phi_j(x)\phi_{\ell-1-j}(x+2\pi\ri n).
\label{eq:tr-int}
\end{align}
To summarize, the computation of trace $\Tr(\rho^\ell W^n)$
boils down to the
construction of a sequence of functions $\phi_j(x)~(j=0,1,\cdots)$.
We should stress that $\phi_j(x)$ is the {\it same function}
appearing in the computation of trace $\Tr\rho^\ell$.
Note that for the integral \eqref{eq:tr-int} to converge, $k$ should satisfy
the condition
\begin{align}
 k>2n.
\end{align}

For some integer values of $k$,
the integral in \eqref{eq:phij} and \eqref{eq:tr-int}
can be evaluated exactly by closing the contour and 
picking up the residue of poles \cite{Okuyama:2011su,PY,HMO1,HMO2}.
For instance, 
we have computed the exact values of $\Tr(\rho^\ell W)$
with the winding number $n=1$,
for $k=3,4,6$ up to certain $\ell$.
For example, for $k=3$ we find
\begin{align}
 \begin{aligned}
  \Tr(\rho W)&=\frac{1}{6}+\ri \frac{1}{2 \sqrt{3}},\\
\Tr(\rho^2 W) &=\frac{ 2-\sqrt{3}}{36}+\ri \frac{ 2-\sqrt{3}}{36}, \\
\Tr(\rho^3 W) &=\frac{9-6 \pi +2 \sqrt{3} \pi }{648 \pi }+\ri 
\frac{9 \sqrt{3}+9 \pi -8 \sqrt{3} \pi }{216 \pi }.
 \end{aligned}
\end{align}
For $k=4$ we find
\begin{align}
 \begin{aligned}
 \Tr(\rho W)&=\frac{1}{8 \sqrt{2}}+\ri \frac{1}{8 \sqrt{2}},\\
\Tr(\rho^2 W)&=\frac{\pi -2}{64 \sqrt{2} \pi }+\ri \frac{4-\pi }{64 \sqrt{2} \pi },\\
\Tr(\rho^3 W)&=\frac{-2-2 \pi +\pi ^2}{512 \sqrt{2} \pi ^2}+\ri
\frac{-3-2 \pi +\pi ^2}{256 \sqrt{2} \pi ^2}.
 \end{aligned}
\label{eq:exact-k4}
\end{align}
For $k=6$ we find
\begin{align}
 \begin{aligned}
  \Tr(\rho W)&=\frac{1}{12 \sqrt{3}}+\ri \frac{1}{36},\\
 \Tr(\rho^2 W)&=\frac{9-\sqrt{3} \pi }{648 \pi }+\ri \frac{7 \pi -12 \sqrt{3}}{432 \pi },\\
 \Tr(\rho^3 W)&=\frac{6-\sqrt{3} \pi }{2592 \pi }+\ri \frac{15 \sqrt{3}-8 \pi }{7776 \pi }.
 \end{aligned}
\end{align}
We note in passing that $\Tr(\rho W^n)$  can be found in a closed form for general $k,n$
\begin{align}
 \Tr(\rho W^n)=\frac{e^{\frac{\pi\ri n^2}{k}}}{4k\cos^2\frac{\pi n}{k}}.
\end{align}
\subsection{Numerical computation}\label{sec:num}
It is useful to 
develop a technique to compute
the trace $\Tr(\rho^\ell W^n)$ numerically with high precision.
We use the observation that the
integral appearing in the computation of $\phi_j(x)$ in \eqref{eq:phij}
is written schematically as
\begin{align}
 \int \frac{dy}{\hbar} \frac{1}{2\cosh\frac{x-y}{2k}}\psi(y)
 =\int \frac{dp}{2\pi} 
 \frac{e^{\frac{\ri px}{\hbar}}}{2\cosh\frac{p}{2}}\til{\psi}(p),
\label{eq:int-sch}
\end{align}
where $\til{\psi}(p)$ is the Fourier transform of $\psi(x)$
\begin{align}
 \til{\psi}(p)=\int \frac{dy}{\hbar} e^{-\frac{\ri py}{\hbar}}\psi(y).
\end{align}
Thus the integral \eqref{eq:int-sch} can be
 evaluated by a successive application of
the Fourier transformation (FT) and the inverse
Fourier transformation $(\text{FT}^{-1})$
\begin{align}
 \int dy \frac{1}{2\cosh\frac{x-y}{2k}}\psi(y)=
 \text{FT}^{-1}\Biggl[\frac{1}{2\cosh\frac{p}{2}}\cdot
 \text{FT}[\psi](p)\Biggr](x).
\label{eq:Fconv}
\end{align}
This can be easily done numerically by
disretizing the continuous variable
$x\in\mathbb{R}$ to some finite number of points in the range 
$-L/2<x\leq L/2$
\begin{align}
 x_i=\left(\frac{i}{m}-\hf\right)L,\qquad (i=1,\cdots,m),
\end{align}
where $L$ is an IR cut-off.
Then we can approximate the Fourier transformation
by the discrete Fourier transformation\footnote{
We compute the trace $\Tr \rho^\ell W^n$ numerically using  a {\tt Mathematica} program
implementing this algorithm, originally written by Yasuyuki Hatsuda.
We are grateful to him for sharing his program with us.
The numerical data of the traces are available upon request to the author.}.
By taking $L$ and $m$ sufficiently large,
we can numerically evaluate the integral \eqref{eq:Fconv}
with very high precision.
We will use the parameters $L=3000$ and $m=2^{16}$ in the numerical computations below. 

\begin{figure}[tb]
\begin{center}
\includegraphics[width=7cm]{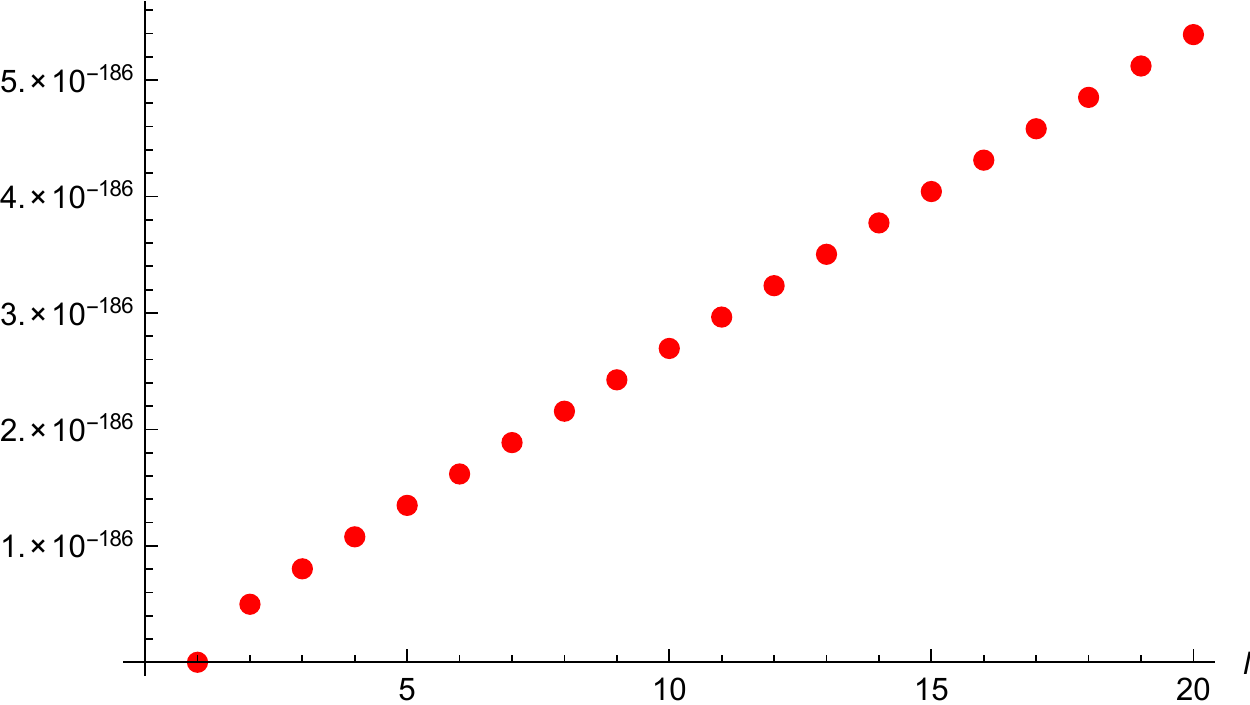}
\end{center}
  \caption{
Plot of 
the numerical error $e_\ell$ \eqref{eq:error-def} of the imaginary part of trace 
$\Tr(\rho^\ell W)$ for $k=4$.
}
  \label{fig:error}
\end{figure}

For integer value of $k$, we can 
estimate the error of the above numerical computation by
comparing with the exact values of traces obtained in the previous subsection.
Let us consider the
imaginary part of $\Tr(\rho^\ell W^n)$ for $k=4,n=1$, as an example.
For $k=4$,
we have computed the exact values of $\Tr(\rho^\ell W)$ up to $\ell=20$
(see \eqref{eq:exact-k4} for the first three terms).
In Fig.~\ref{fig:error}, we show the plot of the relative error $e_\ell$
between the exact values 
and the numerical values of 
the
imaginary part of  $\Tr(\rho^\ell W)$
\begin{align}
 e_{\ell}=\Biggl|\frac{\text{Im}\Tr(\rho^\ell W)_\text{numerical}}{\text{Im}\Tr(\rho^\ell W)_\text{exact}}-1\Biggr|.
\label{eq:error-def}
\end{align}
As we can see from Fig.~\ref{fig:error},
the numerical  error $e_\ell$
is extremely small
\begin{align}
 e_\ell
<6\times 10^{-186},\quad(\ell\leq20).
\label{eq:num-error}
\end{align}
On the other hand, the order of worldsheet 1-instanton correction and
membrane 1-instanton correction for $k=4,N=20$ are
given by (see \eqref{eq:inst-cano})
\begin{align}
\begin{aligned}
\text{worldsheet~1-instanton}:&\quad e^{-2\pi\rt{\frac{2N}{k}}}\approx 2.35\times 10^{-9},\\ 
\text{membrane~1-instanton}:&\quad e^{-2\pi\rt{\frac{kN}{2}}}\approx 5.52\times 10^{-18}. 
\end{aligned}
\label{eq:order-inst}
\end{align}
Comparing \eqref{eq:num-error}
and \eqref{eq:order-inst},
we conclude that our numerical calculation has enough accuracy to study instanton corrections.

\section{Perturbative part of winding Wilson loops}\label{sec:pert}
We will study the large $\mu$ expansion of the
grand canonical VEV of 1/6 BPS winding Wilson loops.
First we consider the perturbative part (or zero-instanton sector)
of this expansion.

In the large $\mu$ limit, the grand canonical VEV
of 1/6 Wilson loops with winding number $n$
behaves  as \cite{KMSS}
\begin{align}
 \bra \Tr U^n\ket^\text{GC}\sim e^{\frac{2n\mu}{k}}.
\end{align}
We would like to understand the large $\mu$
behavior of 1/6 Wilson loops in more detail.
It is shown in \cite{KMSS} that $\bra \Tr U^n\ket^\text{GC}$
can be decomposed as
\begin{align}
 \bra \Tr U^n\ket^\text{GC}=\ri^{n-1}\left(\hf |\cW_n^{(1/2)}|+\ri \cW_n\right).
\label{eq:Un-reim}
\end{align}
The first term is the absolute value of the 1/2 BPS Wilson loop
which was studied extensively in the literature \cite{KMSS,HMMO-Wilson}.
The perturbative part of 1/2 BPS Wilson loop is given by
\cite{KMSS}
\begin{align}
 |\cW_n^{(1/2),\text{pert}}|=\frac{e^{\frac{2n\mu}{k}}}{2\sin\frac{2\pi n}{k}}.
\label{eq:W1/2-pert}
\end{align}
It is known that 
the 1/2 BPS Wilson loops are related to the open topological string
on local $\mathbb{P}^1\times \mathbb{P}^1$ \cite{Marino:2009jd,Grassi:2013qva,HO-1/2,Marino:2016rsq},
and the worldsheet instanton corrections to the 1/2 BPS Wilson loops can be readily obtained
from the results of  open topological string.

On the other hand, $\cW_n$ in \eqref{eq:Un-reim} is less understood. Note that
$\cW_n$ is written as
\begin{align}
 \cW_n=\text{Im}\Bigl[i^{1-n}\bra \Tr U^n\ket^\text{GC}\Bigr].
\label{eq:W-Im}
\end{align}
We will loosely call $\cW_n$  ``the imaginary part of 1/6 BPS Wilson loop''
with the understanding that the precise meaning is \eqref{eq:W-Im}.
In this paper,
we will study the instanton corrections in $\cW_n$.
To do that, first we compute the canonical VEV $\cW_n(N,k)$
numerically with high precision using the algorithm
in section \ref{sec:num}.
Then we can find the 
instanton corrections by numerically fitting
with the expansion
\begin{align}
 \cW_n(N,k)
=e^{A(k)}C(k)^{-\frac{1}{3}}\sum_{j,w}a_{j,w}(-\del_N)^j\text{Ai}\Biggl[C(k)^{-\frac{1}{3}}\Big(N-B(k)-\frac{2n}{k}-w\Big)\Biggr],
\label{eq:Zfund-Airy} 
\end{align}
where $a_{j,w}$ is the coefficient in the instanton expansion
of the grand canonical VEV
\begin{align}
\Xi(\mu,k)\cW_n(\mu,k)=e^{J_\text{pert}(\mu,k)+\frac{2n\mu}{k}}\sum_{j,w}a_{j,w}\mu^j e^{-w\mu}.
\label{eq:a-exp}
\end{align}
Here $J_\text{pert}(\mu,k)$ denotes the perturbative part of the grand potential of ABJM theory
\begin{align}
 J_\text{pert}(\mu,k)=\frac{C(k)\mu^3}{3}+B(k)\mu+A(k),
\end{align}
and the coefficients $C(k), B(k)$ and $A(k)$ are given by
\begin{align}
 \begin{aligned}
  C(k)&=\frac{2}{\pi k^2},\qquad B(k)=\frac{1}{3k}+\frac{k}{24},\\
A(k)&=-\frac{k^2\zeta(3)}{8\pi^2}+4\int_0^\infty dx\frac{x}{e^x-1}\log\left(2\sinh\frac{2\pi x}{k}\right).
 \end{aligned}
\end{align}
It is found \cite{Hanada:2012si,Hatsuda:2014vsa,Hatsuda:2015owa} that  $A(k)$ is a certain resummation of the constant map contribution
of topological string on local $\mathbb{P}^1\times \mathbb{P}^1$.
The expansion \eqref{eq:Zfund-Airy} in the canonical picture
and the grand canonical picture \eqref{eq:a-exp}
are related by the integral transformation
\begin{align}
 \cW_n(N,k)=\int_{\mathcal{C}}\frac{d\mu}{2\pi \ri}e^{J(\mu,k)}\cW_n(\mu,k),
\label{eq:C-int}
\end{align}
where $\mathcal{C}$ is the contour in the $\mu$-plane
from $e^{-\frac{\pi\ri}{3}}\infty$ to $ e^{\frac{\pi\ri}{3}}\infty$,
and $J(\mu,k)$ is known as the modified grand potential
which is related to the grand partition function by \cite{HMO2}
\begin{align}
 \Xi(\mu,k)=\sum_{n\in\mathbb{Z}}e^{J(\mu+2\pi\ri n,k)}.
\label{eq:p-sum}
\end{align}
Note that the original integral in \eqref{eq:2pi-int}
is along the finite segment, but we can extend the contour to 
an infinite path $\mathcal{C}$ in \eqref{eq:C-int}
because of the  summation over the $2\pi\ri$-shift \eqref{eq:p-sum}.
\begin{figure}[tb]
\begin{center}
\begin{tabular}{cc}
\hspace{-4mm}
\includegraphics[width=7cm]{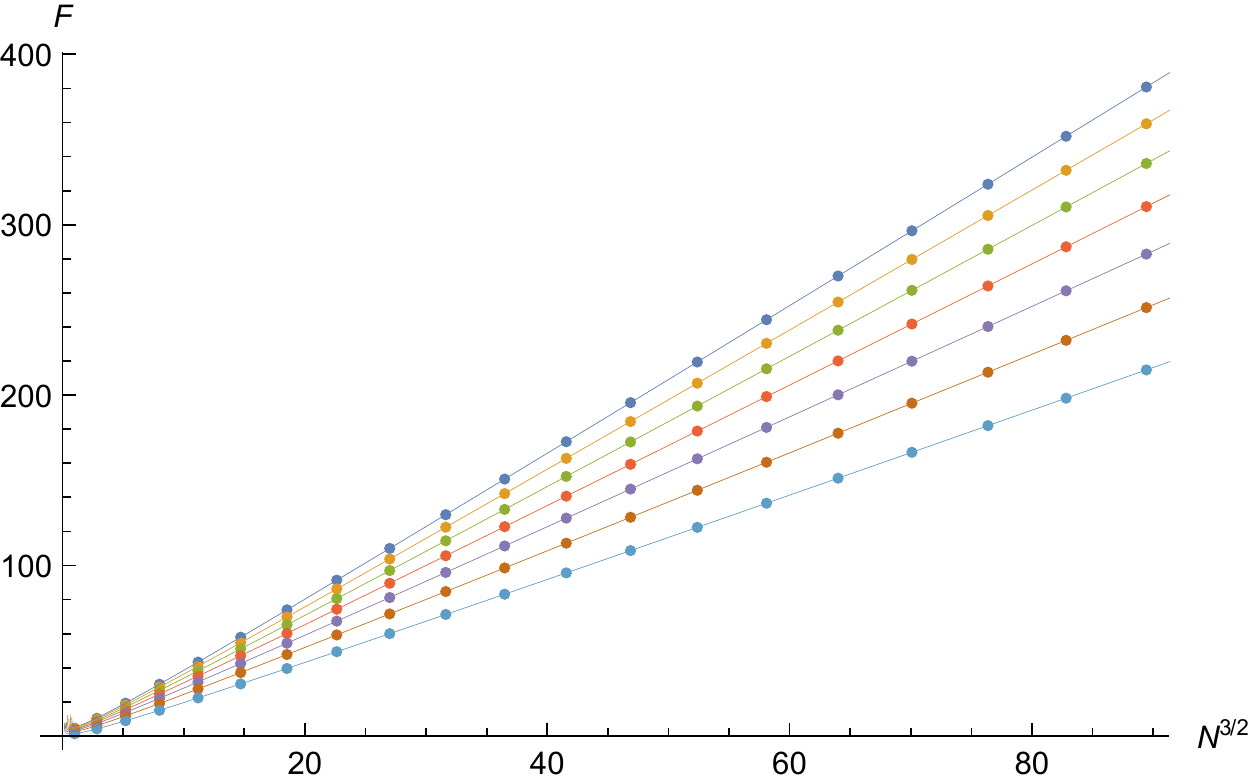}
\hspace{10mm}
&
\includegraphics[width=7cm]{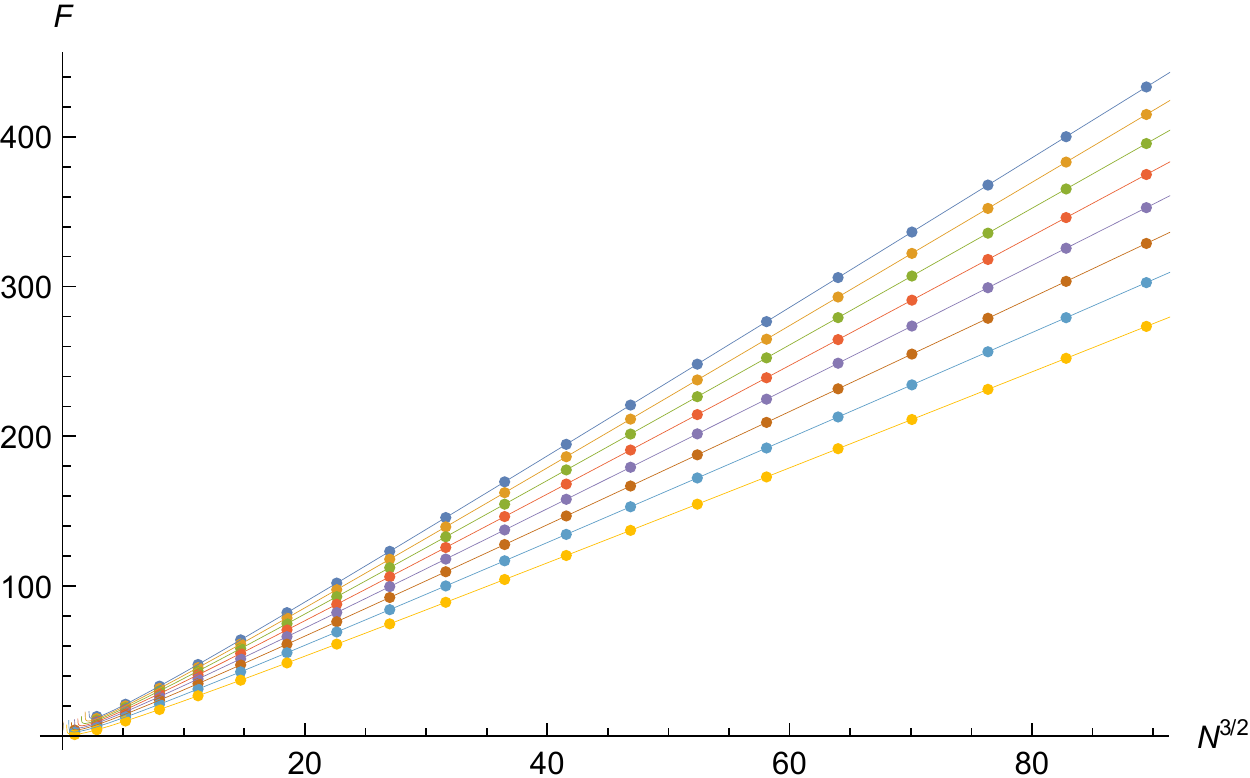}\\
(a) winding number $n=1$
\hspace{10mm}
&
(b) winding number $n=2$
\end{tabular}
\end{center}
  \caption{
This is  the plot 
of ``free energy'' $F=-\log \cW_{n}(N,k)$
for (a) $n=1$ and (b) $n=2$
of the imaginary part of $1/6$ BPS Wilson loop VEV.
Note that the horizontal axis is $N^{3/2}$.
We plot the free energy for $k=3,4,\cdots,9$ in (a) 
and for $k=5,6,\cdots,12$ in (b). $k$ increases from the bottom curve to the top curve
in both (a) and (b).
The dots are the numerical values while the 
solid curves represent the  
perturbative free energy given by the Airy function \eqref{eq:canoWn-pert}.
}
  \label{fig:Zpert-fund}
\end{figure}

From the numerical fitting,
we find that
the perturbative part of the imaginary part of winding Wilson loop is given by
\begin{align}
 \cW_{n}^\text{pert}(\mu,k)=\frac{e^{\frac{2n\mu}{k}}}{k\pi\sin\frac{2\pi n}{k}}
\left(\mu-\pi\sum_{j=1}^n \cot\frac{2\pi j}{k}\right).
\label{eq:Wn-pert}
\end{align}
Using  \eqref{eq:Zfund-Airy}, the corresponding canonical VEV is written as 
\begin{align}
 \cW_{n}^\text{pert}(N,k)=\frac{e^{A(k)}C(k)^{-\frac{1}{3}}}{k\pi\sin\frac{2\pi n}{k}}
\Biggl(-\del_N-\pi\sum_{j=1}^n \cot\frac{2\pi j}{k}\Biggr)\text{Ai}\Biggl[C(k)^{-\frac{1}{3}}\Big(N-B(k)-\frac{2n}{k}\Big)\Biggr].
\label{eq:canoWn-pert}
\end{align}
In Fig.~\ref{fig:Zpert-fund}, we show the plot of ``free energy''
$F=-\log \cW_n(N,k)$ for $n=1,2$ as a function of $N^{3/2}$.
One can clearly see that the numerical value of $\cW_{n}(N,k)$
computed from the algorithm in section \ref{sec:num}
exhibits a nice agreement with the perturbative part given by the Airy function and its derivative \eqref{eq:canoWn-pert}.

\paragraph{Comparison with KMSS.}
In \cite{KMSS} (which we refer to as KMSS), the perturbative part of winding Wilson
loop was obtained as
\begin{align}
\bra \Tr U^n\ket^\text{GC,pert}=\frac{2\pi n e^{\frac{2n\mu}{k}}}{k\sin\frac{2\pi n}{k}}
\Biggl[\Bigl(\mu+\frac{k}{2n}-\pi\cot\frac{2\pi n}{k}\Bigr)A+B\Biggr],
\label{eq:W-KMSS}
\end{align} 
with
\begin{align}
 A=\frac{\ri^n}{2\pi^2n},\qquad B=\ri^{n-1}\frac{k}{4\pi^2n}\Bigl(\frac{\pi}{2}-\ri H_n\Bigr).
\end{align}
Here $H_n$ denotes the harmonic number.
We can recast \eqref{eq:W-KMSS} in the form of
our decomposition in \eqref{eq:Un-reim}
\begin{align}
 \bra \Tr U^n\ket^\text{GC,pert}=\ri^{n-1}\Biggl(\hf|W_n^{(1/2)}|+\ri\cW_n^\text{KMSS}\Biggr).
\end{align}
The first term is the same as \eqref{eq:W1/2-pert}
while the second term is given by
\begin{align}
\cW_n^{\text{KMSS}}=\frac{e^{\frac{2n\mu}{k}}}{k\pi\sin\frac{2\pi n}{k}}\Biggl(\mu-\pi\cot\frac{2\pi n}{k}
-\frac{k}{2}H_{n-1}\Biggr),
\label{eq:Wpert-KMSS}
\end{align}
which is different from our result \eqref{eq:Wn-pert}
for general $n\geq2$.
For the fundamental representation $n=1$, 
\eqref{eq:Wpert-KMSS} agrees with \eqref{eq:Wn-pert} since $H_0=0$.

In the 't Hooft limit
\begin{align}
 k,\mu\to\infty ~~\text{with}~~\frac{\mu}{k}:\text{fixed}
\end{align}
the genus-zero part of \eqref{eq:Wn-pert} is equal to
\eqref{eq:Wpert-KMSS} for general $n$, but the higher genus parts are different.
Note that, for $n\geq2$ the agreement of  \eqref{eq:Wpert-KMSS} 
and the matrix model computation has been  checked in \cite{KMSS} 
only for the genus-zero part\footnote{For the fundamental representation,
the agreement between \eqref{eq:Wpert-KMSS}
and the matrix model was checked for higher genus corrections as well. While for $n>1$,
the authors of \cite{KMSS} used \eqref{eq:Wpert-KMSS}  to predict the higher genus corrections
for general winding. }. 
To see whether our conjecture
\eqref{eq:Wn-pert} is correct or not, we should therefore 
consider the higher genus corrections.
It is likely that the approximation used in \cite{KMSS}
to derive \eqref{eq:Wpert-KMSS} misses some $1/k$ corrections\footnote{We would like to thank Marcos Mari\~no for raising this possibility.}.
Our numerical study strongly suggests that \eqref{eq:Wn-pert} is the correct perturbative part.

\section{Fundamental Wilson loop}\label{sec:fund}
In this section, we will consider the instanton corrections of the imaginary part of 1/6 BPS Wilson loop
$\cW_{1}(\mu,k)$ in the fundamental representation, i.e. the winding number $n=1$.
Using the instanton expansion \eqref{eq:Zfund-Airy}
in terms of the Airy function and its derivatives,
one can find the coefficient $a_{j,w}$ in \eqref{eq:Zfund-Airy}
by fitting the value of canonical VEV $\cW_{1}(N,k)$
computed either exactly (see section \ref{sec:exact}) or numerically (see section \ref{sec:num}).
The expansion coefficients $a_{j,w}$ becomes simple numbers
for some integer $k$, and one can guess 
the exact value of the coefficients from the  numerical fitting. 
The results are summarized in Appendix \ref{app:fund}.
We have also computed the instanton expansion
for fractional $k$ using the numerical computation of the trace 
$\Tr(\rho^\ell W)$ explained in section \ref{sec:num}.
It turns out that the grand canonical VEV of the imaginary part of 1/6 BPS Wilson loop
in the fundamental representation has the following expansion
\begin{align}
 \cW_{1}(\mu,k)=\cW_{1}^\text{pert}(\mu,k)+\sum_{\ell,m}\cW^{(\ell,m)}_{1}(\mu,k)
\end{align}
where $\cW^{(\ell,m)}_{1}(\mu,k)$ denotes the contribution of
the bound state of  membrane $\ell$-instanton and  worldsheet $m$-instanton
 \begin{align}
  e^{-\frac{2\mu}{k}}\cW^{(\ell,m)}_1(\mu,k)\propto e^{-2\ell\mu-\frac{4\mu m}{k}}.
 \end{align}
\paragraph{Worldsheet instantons.}
First, let us consider the worldsheet instanton corrections
(and the perturbative part)
\begin{align}
 \cW_1^{\text{WS}}(\mu,k)=\cW_1^{\text{pert}}(\mu,k)+
 \sum_{m=1}^\infty \cW^{(0,m)}_1(\mu,k).
\end{align}
From the numerical fitting, we find that the worldsheet instanton corrections are given by
\begin{align}
\begin{aligned}
 &e^{-\frac{2\mu}{k}}\cW_1^{\text{WS}}(\mu,k)\\
 =&\frac{\mu}{k\pi\sin\frac{2\pi}{k}}
 \Big[1+2e^{-\frac{4\mu}{k}}-e^{-\frac{8\mu}{k}}+2e^{-\frac{12\mu}{k}}
 -7e^{-\frac{16\mu}{k}}+
\mathcal{O}(e^{-\frac{20\mu}{k}})\Big]\\
 -&\frac{\cos\frac{2\pi}{k}}{k\sin^2\frac{2\pi}{k}}\left[1+0\cdot e^{-\frac{4\mu}{k}}
 -\left(3+\frac{1}{\cos^2\frac{2\pi}{k}}\right)e^{-\frac{8\mu}{k}}\right.\\
 &\hskip20mm +\frac{2\sin\frac{2\pi}{k}}{\sin\frac{6\pi}{k}}
 \left(32\cos^2\frac{2\pi}{k}-10+\frac{1}{\cos^2\frac{2\pi}{k}}\right)e^{-\frac{12\mu}{k}}\\
 &\left. \hskip20mm+\left(\frac{8\sin\frac{2\pi}{k}}{\sin\frac{6\pi}{k}}-\frac{7}{2\cos^2\frac{2\pi}{k}}-\frac{1}{\cos\frac{4\pi}{k}}
 -43-32\cos\frac{4\pi}{k}\right)e^{-\frac{16\mu}{k}} +\mathcal{O}(e^{-\frac{20\mu}{k}})\right].
 \label{eq:WS}
\end{aligned}
\end{align}
Note that the worldsheet instanton in \eqref{eq:WS}
has poles at $k=4,6$, which should be canceled by
the membrane instanton and the bound state.
Thus, we expect that there are membrane instanton corrections
in the imaginary part of 1/6 BPS Wilson loop.
This is in contrast to the 1/2 BPS Wilson loops where
the ``pure'' membrane instanton corrections are absent
except for the bound state corrections coming from
the quantum mirror map  $\mu\to\mu_\text{eff}$ \cite{HMMO-Wilson,HO-1/2}.

\paragraph{Membrane instantons and Bound states.}
From the numerical fitting, we find that the membrane 1-instanton and the $(1,1)$
bound state are given by
\begin{align}
\begin{aligned} 
 e^{-\frac{2\mu}{k}}\cW^{(1,0)}_1&=e^{-2\mu} 
\frac{\cos\frac{\pi k}{2}}{k\pi\sin\frac{2\pi}{k}}
\left[\left(4-\frac{4}{k}\right)\left(\mu-\pi\cot\frac{2\pi}{k}\right)-2-\pi k\cot\frac{\pi k}{2}\right],\\
e^{-\frac{2\mu}{k}}\cW^{(1,1)}_1&=e^{-\frac{4\mu}{k}-2\mu} 
\frac{\cos\frac{\pi k}{2}}{k\pi\sin\frac{2\pi}{k}}\left[\left(8+\frac{8}{k}\right)\mu
-4-2\pi k\cot\frac{\pi k}{2}\right].
\end{aligned}
\label{eq:mem-1}
\end{align}
As expected, one can show that
the poles at $k=4,6$ are canceled between  worldsheet instantons \eqref{eq:WS} 
and  membrane instantons or bound states \eqref{eq:mem-1}, and the remaining finite part reproduces the  result in
Appendix \ref{app:fund}
\begin{align}
 \begin{aligned}
  \lim_{k\to4}e^{-\frac{2\mu}{k}}\Bigl[\cW_1^{(1,0)}+\cW_{1}^{(0,2)}\Bigr]&=\frac{3\mu-1}{2\pi}e^{-2\mu},\\
\lim_{k\to4}e^{-\frac{2\mu}{k}}\Bigl[\cW_1^{(1,1)}+\cW_{1}^{(0,3)}\Bigr]&=\frac{5\mu-1}{\pi}e^{-3\mu} ,\\
\lim_{k\to6}e^{-\frac{2\mu}{k}}\Bigl[\cW_{1}^{(1,0)}+\cW_{1}^{(0,3)}\Bigr]&=\left[\frac{2(-8\mu+3)}{9\rt{3}\pi}-\frac{20}{27}\right]e^{-2\mu},\\
   \lim_{k\to6}e^{-\frac{2\mu}{k}}\Bigl[\cW_1^{(1,1)}+\cW_{1}^{(0,4)}\Bigr]&=
   \left[\frac{-73\mu+12}{9\rt{3}\pi}+\frac{35}{9}\right]e^{-\frac{8\mu}{3}}.
  \end{aligned}
 \end{align}
This pole cancellation mechanism was originally found in the grand potential
of ABJM theory \cite{HMO2}. Here we 
find that the similar pole cancellation mechanism works also in the VEV of 1/6 BPS Wilson loops.

\begin{figure}[tb]
\begin{center}
\begin{tabular}{cc}
\hspace{-4mm}
\includegraphics[width=6cm]{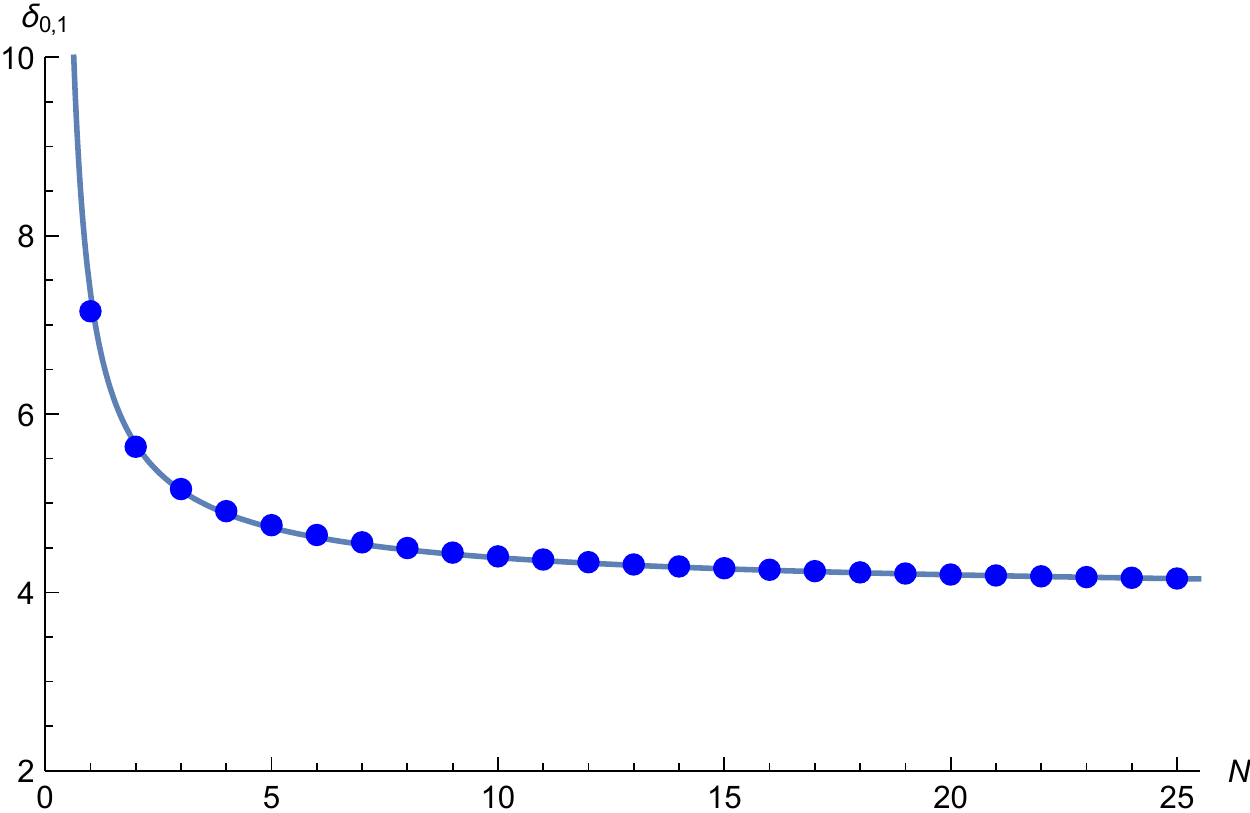}
\hspace{10mm}
&
\includegraphics[width=6cm]{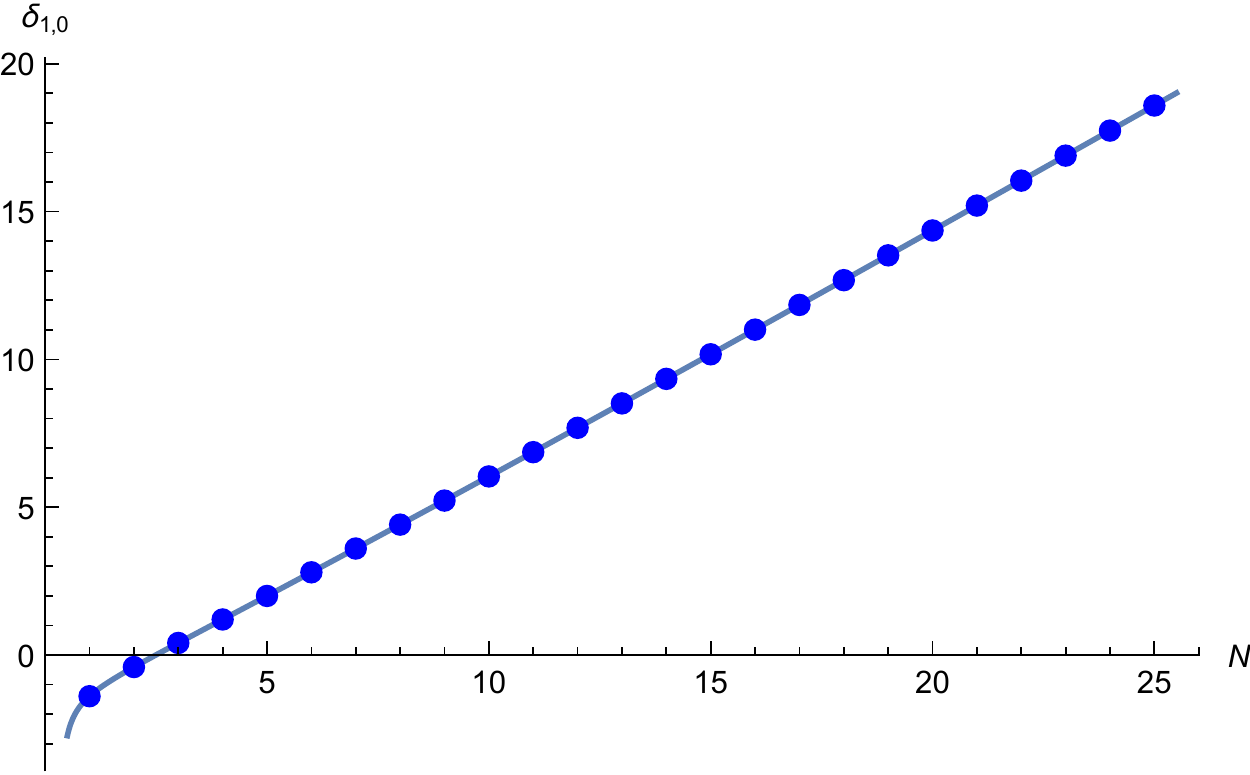}\\
(a) worldsheet 1-instanton
\hspace{10mm}
&
(b) membrane 1-instanton\\
\includegraphics[width=6cm]{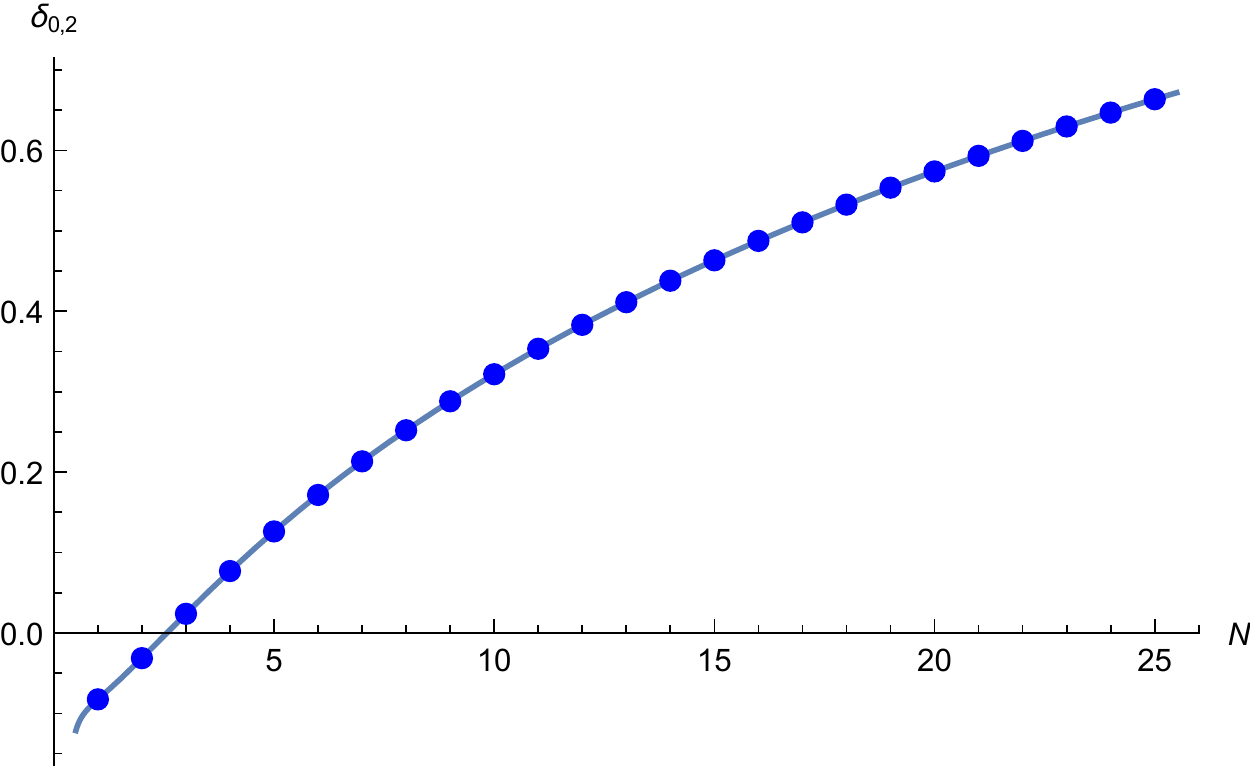}
\hspace{10mm}
&
\includegraphics[width=6cm]{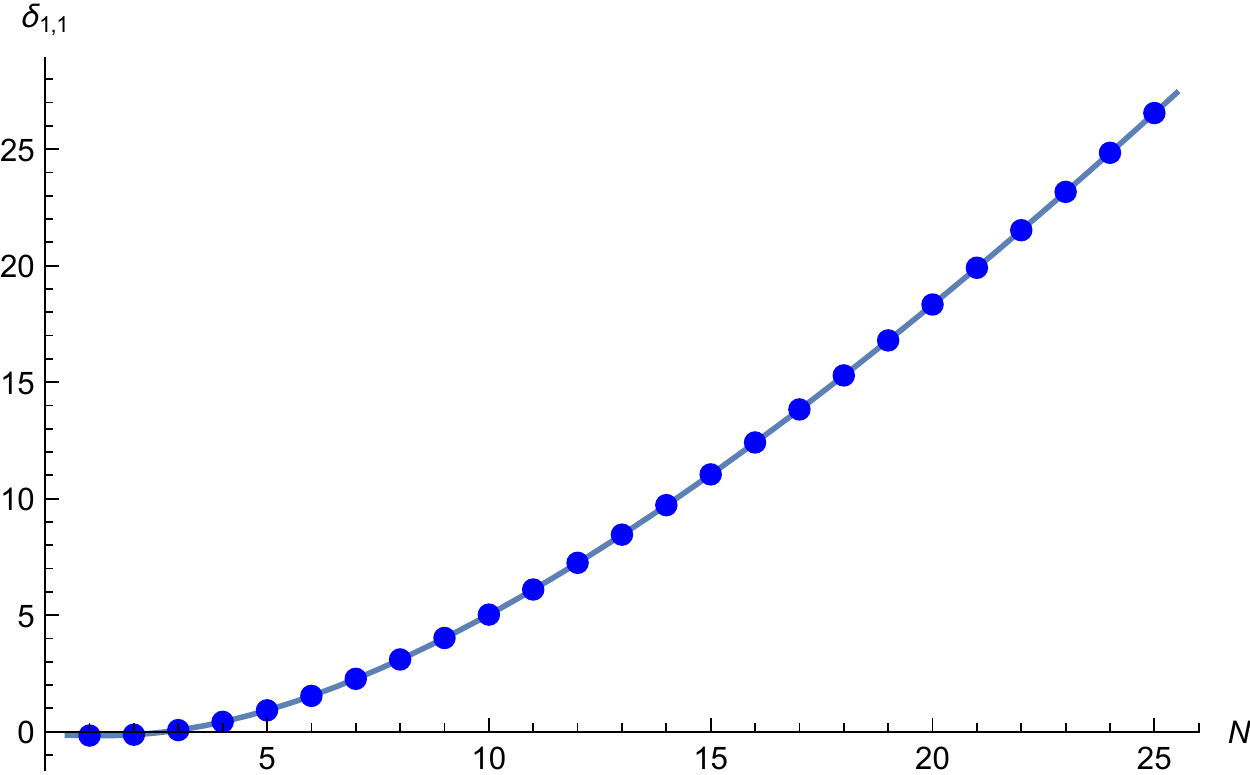}\\
(c) worldsheet 2-instanton
\hspace{10mm}
&
(d) $(1,1)$ bound state
\end{tabular}
\end{center}
  \caption{
This is  the plot 
of 
 (a) $\cob_{0,1}$ (b) $\cob_{1,0}$ (c) $\cob_{0,2}$ and (d) $\cob_{1,1}$ against $N$,
for $k=2.7$.
The dots are the numerical values of \eqref{eq:num-cob} 
while the solid curves represent the  
expected behavior of $(\ell,m)$ instanton given by the Airy function and its derivatives \eqref{eq:Airy-cob}.
}
  \label{fig:bound}
\end{figure}

Let us see that our conjecture of instanton corrections \eqref{eq:WS}
and \eqref{eq:mem-1} 
correctly reproduces the behavior of $\cW_1(N,k)$  for non-integer $k$ as well.
We take $k=2.7$ as an example. 
For $k=2.7$, the instanton factors have the following ordering
\begin{align}
 e^{-\frac{4\mu}{k}}>e^{-2\mu}>e^{-\frac{8\mu}{k}}>e^{-\frac{4\mu}{k}-2\mu}>\cdots.
\label{eq:order}
\end{align}
Once we know the instanton correction in the grand canonical picture $\cW_1^{(\ell,m)}(\mu,k)$, 
we can easily translate it to the canonical picture 
$\cW_1^{(\ell,m)}(N,k)$ using \eqref{eq:Zfund-Airy}, and as a consequence
the instanton correction $\cW_1^{(\ell,m)}(N,k)$ can be written as
an Airy function and its derivatives.
From \eqref{eq:order}, we define the 
the quantity $\cob_{\ell,m}$ by subtracting the instanton corrections
up to the order $e^{-2\ell\mu-\frac{4m\mu}{k}}$
(we do not subtract the term $\cW_1^{(\ell,m)}(N,k)$
in defining $\cob_{\ell,m}$)
\begin{align}
\begin{aligned}
\cob_{0,1}&=\frac{\cW_1(N,k)-\cW_1^\text{pert}(N,k)}{\cW_1^\text{pert}(N,k)}e^{\frac{4\mu_*}{k}},\\
 \cob_{1,0}&=\frac{\cW_1(N,k)-\cW_1^\text{pert}(N,k)-\cW^{(0,1)}_1(N,k)}{\cW_1^\text{pert}(N,k)}e^{2\mu_*},\\
\cob_{0,2}&=\frac{\cW_1(N,k)-\cW_1^\text{pert}(N,k)-\cW_1^{(0,1)}(N,k)-\cW_1^{(1,0)}(N,k)}{\cW_1^\text{pert}(N,k)}e^{\frac{8\mu_*}{k}},\\
 \cob_{1,1}&=\frac{\cW_1(N,k)-\cW_1^\text{pert}(N,k)-\cW_1^{(0,1)}(N,k)-\cW_1^{(1,0)}(N,k)-\cW_1^{(0,2)}(N,k)}{\cW_1^\text{pert}(N,k)}e^{2\mu_*+\frac{4\mu_*}{k}}. 
\end{aligned}
\label{eq:num-cob}
\end{align}
We have included the exponential factor $e^{2\ell\mu_*+\frac{4m\mu_*}{k}}$
in the definition of $\cob_{\ell,m}$ 
with $\mu_*$
being the saddle point value of the chemical potential\footnote{
Note that this relation \eqref{eq:mu-*} implies
that the worldsheet instanton and the membrane instanton
factors in the canonical picture are given by
\begin{align}
 e^{-\frac{4\mu_*}{k}}=e^{-2\pi \rt{\frac{2N}{k}}},\quad
e^{-2\mu_*}=e^{-2\pi \rt{\frac{kN}{2}}}.
\label{eq:inst-cano}
\end{align}
We have already used
this relation  in \eqref{eq:order-inst}.}
\begin{align}
 \mu_*=\rt{\frac{N}{C(k)}}=\pi\rt{\frac{kN}{2}}.
\label{eq:mu-*}
\end{align}
The canonical VEV $\cW_1(N,k)$
in \eqref{eq:num-cob} can be evaluated numerically with high precision
using the method in section \ref{sec:num} even for a non-integer value of $k=2.7$.
We expect that $\cob_{\ell,m}$ is approximated by
\begin{align}
 \begin{aligned}
 \cob_{\ell,m}&\approx \frac{\cW_1^{(\ell,m)}(N,k)}{\cW_1^\text{pert}(N,k)}e^{2\ell\mu_*+\frac{4m\mu_*}{k}}. 
 \end{aligned} 
\label{eq:Airy-cob}
\end{align}
As shown in Fig.~\ref{fig:bound},
we find a nice agreement between 
\eqref{eq:num-cob} and \eqref{eq:Airy-cob}, as expected.
This confirms the validity of our conjecture
of instanton corrections \eqref{eq:WS} and \eqref{eq:mem-1}, for $k=2.7$.
We have performed similar checks for various values of $k$.

\paragraph{Rewriting in terms of $\mu_\text{eff}$.}
Let us rewrite $\cW_1(\mu,k)$ in terms of the ``effective''
chemical potential $\mu_\text{eff}$, which was first introduced in \cite{HMO2}
\begin{align}
 \mu_\text{eff}=\mu-2\cos\frac{\pi k}{2}e^{-2\mu}-(4+5\cos\pi k)e^{-4\mu}+\cdots.
\end{align}
This is interpreted as the quantum mirror map of local
$\mathbb{P}^1\times \mathbb{P}^1$ \cite{HMMO}
and the coefficient of this expansion can be easily obtained
from the formal solution of the wavefunction annihilated by the quantized mirror
curve of local
$\mathbb{P}^1\times \mathbb{P}^1$ \cite{Aganagic:2011mi}.

After rewriting the instanton corrections in \eqref{eq:WS}
and \eqref{eq:mem-1}
in terms of $\mu_\text{eff}$,
somewhat miraculously, we find that
$\cW_1(\mu,k)$ is completely factorized up to the $(1,1)$ bound state 
\begin{align}
 \cW_1=\frac{e^{\frac{2\mu_\text{eff}}{k}}}{k\pi\sin\frac{2\pi}{k}}
(1+2Q_w)(1+4\cos \frac{\pi k}{2}Q_m)
\left[\mu_\text{eff}-\pi\cot\frac{2\pi}{k}(1-2Q_w)-\pi k \cos \frac{\pi k}{2}\cot \frac{\pi k}{2}Q_m\right],
\end{align}
where we have introduced the worldsheet instanton factor 
$Q_w$ and the membrane instanton factor $Q_m$ by
\begin{align}
 Q_w=e^{-\frac{4\mu_\text{eff}}{k}},\qquad
Q_m=e^{-2\mu_\text{eff}}.
\end{align}

Assuming that this factorized structure holds for higher instanton numbers,
we can continue the numerical fitting of instanton coefficients.
In this way, we find that $\cW_1$
is written as
\begin{align}
 \cW_1=\frac{e^{\frac{2\mu_\text{eff}}{k}}}{k\pi\sin\frac{2\pi}{k}}
f_wf_m\Big(\mu_\text{eff}-\pi \cot \frac{2\pi}{k} V_w-\pi k V_m \Big),
\label{eq:fact-fund}
\end{align}
where
\begin{align}
 \begin{aligned}
  f_w&=1+2Q_w-Q_w^2+2Q_w^3-7Q_w^4+16\cos^2\frac{2\pi}{k}\Bigl(3-\cos\frac{4\pi}{k}\Bigr)Q_w^5+\\
&+\Bigl(-72 \cos \left(\frac{4 \pi }{k}\right)-24 \cos \left(\frac{8
   \pi }{k}\right)+8 \cos \left(\frac{12 \pi }{k}\right)+6
   \cos \left(\frac{16 \pi }{k}\right)-90\Bigr)Q_w^6+\mathcal{O}(Q_w^7),
\end{aligned}
\end{align}
\begin{align}
 \begin{aligned}
V_w&=1-2Q_w+\frac{\cos\frac{4\pi}{k}}{\cos^2\frac{2\pi}{k}}Q_w^2
   +4\left(2+\frac{\sin\frac{2\pi}{k}}{\sin\frac{6\pi}{k}}\right)
   Q_w^3-2\left(23+\frac{\left(1+4\cos\frac{4\pi}{k}\right)^2\sin\frac{6\pi}{k}}{\cos\frac{2\pi}{k}\sin\frac{8\pi}{k}}\right)Q_w^4\\
&+\left(322+152 \cos \left(\frac{4 \pi }{k}\right)+108 \cos
   \left(\frac{8 \pi }{k}\right)+\frac{8 \sin \left(\frac{2 \pi
   }{k}\right) \cos \left(\frac{4 \pi }{k}\right)}{ \sin
  \left(\frac{10 \pi }{k}\right)}\right)Q_w^5\\
  &+\Biggl(-1428-1416 \cos \left(\frac{4 \pi }{k}\right)-816 \cos
   \left(\frac{8 \pi }{k}\right)-248 \cos \left(\frac{12 \pi
   }{k}\right)-84 \cos \left(\frac{16 \pi }{k}\right)\\
  &\hskip10mm-\frac{82}{3\cos ^2\left(\frac{2 \pi
  }{k}\right)} +\frac{4}{3} \frac{\left(13 \sin \left(\frac{4 \pi
   }{k}\right)-14 \sin \left(\frac{8 \pi }{k}\right)\right)}{
  \sin \left(\frac{12 \pi }{k}\right)}\Biggr)Q_w^6+\mathcal{O}(Q_w^7),\\
 \end{aligned}
\label{eq:WIm-1}
\end{align}
and
\begin{align}
 \begin{aligned}
f_m&=1+4\cos \frac{\pi k}{2}Q_m+4(2+3\cos\pi k)Q_m^2+8\cos \frac{\pi k}{2}(5+9\cos\pi k+3\cos2\pi k)Q_m^3
+\mathcal{O}(Q_m^4),\\
V_m&=\cos \frac{\pi k}{2}\cot \frac{\pi k}{2}Q_m
+(4+5\cos\pi k)\cot\pi k Q_m^2+
2\cot\frac{3\pi k}{2} \cos \frac{\pi k}{2} (13+19\cos\pi k+9\cos2\pi k)Q_m^3
+\mathcal{O}(Q_m^4).
\label{eq:fm-Vm}
 \end{aligned}
\end{align}
We observe that the coefficient of $V_m$ is proportional to the membrane instanton coefficient
$\til{b}_\ell(k)$ in the modified grand potential of ABJM theory \cite{HMO2}
\begin{align}
 J(\mu,k)=J_\text{WS}(\mu_\text{eff})+\sum_{\ell=1}^\infty \Biggl[\mu_{\text{eff}}\til{b}_\ell(k)
-k^2\frac{\del}{\del k}\Biggl(\frac{\til{b}_\ell(k)}{2\ell k}\Biggr)\Biggr]e^{-2\ell\mu_\text{eff}}.
\end{align}
From this observation, we conjecture
that $V_m$ is given by
\begin{align}
 V_m=-\frac{\pi}{4}\frac{\del\til{J}_b}{\del\mu_\text{eff}},\qquad
\til{J}_b=\sum_{\ell=1}^\infty \til{b}_\ell(k)e^{-2\ell\mu_\text{eff}}.
\label{eq:Vm-Jb}
\end{align}
We also observe that $f_m$ in \eqref{eq:fm-Vm}
is equal to the derivative of the effective chemical potential
\begin{align}
 f_m=\frac{\del\mu_\text{eff}}{\del\mu}.
\end{align}

\paragraph{Relation to quantum volume.}
As shown in \cite{HMMO}, the membrane instanton part
$\til{J}_b$ of the modified grand potential of ABJM theory is
given by the NS free energy
on local $\mathbb{P}^1\times \mathbb{P}^1$.
From this relation, $V_m$ in \eqref{eq:Vm-Jb} is further rewritten as\footnote{For a general local Calabi-Yau, the NS free energy is written as
\begin{align}
 F_\text{NS}(\boldsymbol{Q};e^{\ri\hbar})=\sum_{j_L,j_R,\boldsymbol{d}}
N_{j_L,j_R}^{\boldsymbol{d}}\sum_{w=1}^\infty
\frac{\sin\frac{\hbar w}{2}(2j_L+1)\sin\frac{\hbar w}{2}(2j_R+1)}{2w^2\sin^3\frac{\hbar w}{2}}
\prod_I Q_I^{wd_I}.
\end{align}
}
\begin{align}
 V_m=-\hf\frac{\del^2}{\del t^2}F_\text{NS}(Q_mq^{\hf},Q_mq^{-\hf};q),
\label{eq:Vm-FNS}
\end{align}
where we defined
\begin{align}
 Q_m=e^{-t},\quad t=2\mu_{\text{eff}},\quad q=e^{\ri\hbar_\text{top} },\quad
\hbar_\text{top}=\pi k.
\end{align}
Note that the normalization of the Planck constant
$\hbar_\text{top}$ in topological string is different 
from  the $\hbar$ of  ABJM Fermi gas \eqref{eq:hbar} by a factor of $2$, which comes from
rewriting the operator $\rho^{-1}$ into the canonical form of
the mirror curve of  local $\mathbb{P}^1\times \mathbb{P}^1$
\cite{Kallen:2013qla}.
Appearance of the NS free energy in \eqref{eq:Vm-FNS}
suggests that the imaginary part of 1/6 BPS Wilson loop
is closely related to the quantum volume $\Om$ of the phase space, 
which determines the exact quantization condition
of the operator $\rho$
\begin{align}
 \Om=2\pi\left(n+\hf\right).
\end{align}
It is found in \cite{Kallen:2013qla,Wang:2014ega,Wang:2015wdy,Hatsuda:2015fxa}
that the quantum volume $\Om$ has two pieces
\begin{align}
 \Om=\Om^\text{WKB}+\Om^\text{np},
\end{align}
where the WKB part $\Om^\text{WKB}$
is given by the NS free energy
\begin{align}
\begin{aligned}
 \Om^\text{WKB}&=\frac{t^2}{\hbar_\text{top}}
-\frac{2\pi^2}{3\hbar_\text{top}}+\frac{\hbar_\text{top}}{12}+2\frac{\del}{\del t} 
F_\text{NS}(Qq^{\hf},Qq^{-\hf};q),
\end{aligned}
\label{eq:Om-WKB}
\end{align}
while the non-perturbative part $\Om^\text{np}$
of quantum volume 
is given by the ``S-dual'' of 
$\Om^\text{WKB}$
\begin{align}
\Om^\text{np}&=2\frac{\del}{\del t_D} 
F_\text{NS}(-Q_w,-Q_w;q_D),  
\label{eq:Om-np}
\end{align}
where the ``S-dual'' variables are given by
\begin{align}
 Q_w=e^{-t_D},\quad t_D=\frac{2\pi}{\hbar_\text{top}}t,\quad
q_D=e^{\ri \hbar_{\text{top}}^D},\quad
\hbar_{\text{top}}^D=\frac{4\pi^2}{\hbar_{\text{top}}}.
\end{align}
We find that the membrane instanton part of $\cW_1(\mu,k)$ is 
related to $\Om^\text{WKB}$ by
\begin{align}
 \mu_\text{eff}-\pi k V_m=\frac{\pi k}{4}\del_t\Om^\text{WKB}.
\end{align}

It is found that the pole cancellation at rational value of $k$
is guaranteed by the combination of $\Om^\text{WKB}+\Om^\text{np}$.
Since the pole cancellation also occurs in the imaginary part of 1/6 BPS Wilson loop
$\cW_1(\mu,k)$, it is natural to expect that the worldsheet instanton correction
$V_w$ in  \eqref{eq:WIm-1} is related to the S-dual of NS free energy in \eqref{eq:Om-np}.
In fact, 
if we define
\begin{align}
 \del_{t_D}\til{\Om}^\text{np}=-2\cot\frac{2\pi}{k}V_w,
\end{align}
we find that the $\del_{t_D}\til{\Om}^\text{np}$ and
$\del_{t_D}\Om^{\text{np}}$ have the same singularity structure,
and their difference is regular in the convergence region $k>2$ of $\cW_1(\mu,k)$
\begin{align}
  \del_{t_D}(\til{\Om}^\text{np}-\Om^{\text{np}})=2\cot\frac{2\pi}{k}
\Biggl[-1+8\cos\frac{4\pi}{k}Q_w^2-4\Bigl(7+9\cos\frac{8\pi}{k}\Bigr)Q_w^3+\cdots\Biggr].
\label{eq:reg}
\end{align}
Putting all together, we arrive at a surprisingly simple expression
of the imaginary part of the 1/6 BPS Wilson loop in the fundamental representation
\begin{align}
 \cW_1(\mu,k)=\frac{e^{\frac{2\mu_\text{eff}}{k}}f_w}{4\sin\frac{2\pi}{k}}
\frac{\del t}{\del(2\mu)}\frac{\del}{\del t}(\Om^\text{WKB}+\til{\Om}^\text{np})
=\frac{e^{\frac{2\mu_\text{eff}}{k}}f_w}{8\sin\frac{2\pi}{k}}
 \frac{\del}{\del \mu}(\Om^\text{WKB}+\til{\Om}^\text{np}).
 \label{eq:W1-final}
\end{align}
Currently we do not have a clear understanding of the
physical meaning of the regular part in \eqref{eq:reg}.
The factor $f_w$ might be interpreted as a part of the definition of
the ``open flat coordinate'' \cite{Aganagic:2001nx}
for the 1/6 BPS Wilson loops.
It would be very interesting to understand the deep reason of the connection between
the 1/6 BPS Wilson loops and the quantum volume. 

\subsection{Genus expansion of 1/6 BPS Wilson loops at large $\la$}
Using \eqref{eq:WS}, we can predict the 't Hooft expansion of 1/6 BPS Wilson loop
at large 't Hoot coupling $\la=N/k$.
Let us consider the genus expansion of the canonical VEV of 
the imaginary part of 1/6 BPS Wilson loop in the fundamental representation,
normalized by the partition function
\begin{align}
 \frac{\cW_{1}(N,k)}{Z(N,k)}=\frac{e^{\hf s}}{\pi \ri}\sum_{g=0}^\infty g_s^{2g-1}\,
\cW_{1}^{(g)},
\end{align}
where the string coupling $g_s$ and the parameter $s$ are defined by
\begin{align}
 g_s=\frac{4\pi\ri}{k},\qquad
s=2\pi \rt{2\h{\la}},\qquad \h{\la}=\la-\frac{1}{24}.
\end{align}
From the planar solution of the resolvent of ABJM matrix model,
the genus-zero part of the fundamental Wilson loop is written as \cite{Marino:2009jd,Drukker:2010nc}
\begin{align}
 \lim_{N,k\to\infty}\frac{\frac{1}{N}\bra\Tr U\ket_{N}}{Z(N,k)}
=\frac{1}{2\pi^2\ri \la}\int_{-a}^a\, dx e^{x}\arctan\rt{\frac{\sinh^2 \frac{a}{2}-\sinh^2 \frac{x}{2}}{\cosh^2 \frac{x}{2}-\sinh^2 \frac{a}{2}}},
\end{align}
where the end-point of the cut $a$ is given by
\begin{align}
\ri\ka=4\sinh^2\frac{a}{2},
\end{align}
and $\la$ and $\ka$ are related by
\begin{align}
 \la=\frac{\ka}{8\pi}{}_3F_2\left(\hf,\hf,\hf;1,\frac{3}{2};-\frac{\ka^2}{16}\right).
\end{align}
We find that the imaginary part of 1/6 BPS Wilson loop at genus-zero 
can be written in a simple form
\begin{align}
\begin{aligned}
 \cW_1^{(g=0)}&=-\pi^2\int d\ka \,\ka\frac{d\la}{d\ka}=4E\left(-\frac{\ka^2}{16}\right)
-\frac{(\ka^2+16)}{4}K\left(-\frac{\ka^2}{16}\right),
\end{aligned}
\end{align}
where $K(k^2)$ and $E(k^2)$ denote
the complete elliptic integrals of the first and the second kinds, respectively.

The worldsheet instanton corrections of $\cW_1(\mu,k)$ in  \eqref{eq:WS}
in the grand canonical picture can be easily translated to
the canonical picture in the 't Hooft limit
\begin{align}
 \begin{aligned}
  \cW_1^{(g=0)}&=1-\frac{s}{2}-(1 + s)e^{-s}+\left(\frac{s}{2}-\frac{1}{s}+\frac{1}{4}\right)e^{-2s}
+\left(\frac{4}{3 s^3}+\frac{10}{3
   s^2}-s+\frac{5}{2
   s}-\frac{23}{18}\right)e^{-3s}+\mathcal{O}(e^{-4s}),\\
\cW_1^{(g=1)}&=-\frac{1}{24}\left[1-\frac{s}{2}-\left(1+s\right)e^{-s}
+\left(\frac{s}{2}-\frac{1}{s}+\frac{1}{4}\right)e^{-2s}
+\left(\frac{4}{3 s^3}+\frac{10}{3
   s^2}-s+\frac{5}{2
   s}-\frac{23}{18}\right)e^{-3s}+\mathcal{O}(e^{-4s})\right],
\\
\cW_1^{(g=2)}&=\frac{5}{192 s^3}-\frac{7}{384 s^2}-\frac{7 s}{11520}+\frac{1}{144
   s}-\frac{1}{5760}\\
&+\left(-\frac{5}{16 s^5}-\frac{11}{96 s^4}-\frac{43}{576 s^3}+\frac{19}{576
   s^2}-\frac{7 s}{5760}+\frac{11}{1440 s}-\frac{31}{5760}\right)e^{-s}\\
&+\left(\frac{25}{8 s^7}+\frac{209}{48 s^6}+\frac{719}{192
   s^5}+\frac{755}{384 s^4}+\frac{8629}{11520 s^3}-\frac{169}{1920
   s^2}+\frac{7 s}{11520}-\frac{1483}{5760 s}+\frac{47}{2560}\right)e^{-2s}+\mathcal{O}(e^{-3s}).
 \end{aligned}
\label{eq:genus}
\end{align}
One can show that the genus-zero and the genus-one
amplitudes in \eqref{eq:genus} are consistent with
the matrix model results \cite{Drukker:2010nc}.
Interestingly, from \eqref{eq:genus}
we observe that the genus-zero and the genus-one amplitudes are proportional to each other
\begin{align}
 \cW_1^{(g=1)}=-\frac{1}{24}\cW_1^{(g=0)}.
\end{align}
It would be interesting to prove this relation directly from the matrix model.

\section{WKB expansion}\label{sec:WKB}
In this section, we study the membrane instanton corrections of 1/6 BPS Wilson loops 
in the fundamental representation from 
the WKB expansion of Fermi gas.
As discussed in \cite{Hatsuda:2015oaa},
the WKB expansion can be systematically analyzed
using the spectral trace and the Mellin-Barnes representation. 

First, we rewrite the small $\ka$ expansion
of $\bra \Tr U\ket^{\text{GC}}$ in \eqref{eq:exp-trU}
into the Mellin-Barnes type integral representation
\begin{align}
 \bra \Tr U\ket^{\text{GC}}=\int_{c-\ri\infty}^{c+\ri\infty}\frac{ds}{2\pi i}
\frac{\pi}{\sin\pi s}\Tr(\rho^sW)e^{s\mu},
\end{align}
where $c$ is a constant in the range $2/k<c<1$.
Picking up the pole at $s=\ell\in\mathbb{N}$, we recover the small
$\ka$ expansion \eqref{eq:exp-trU}.
On the other hand, closing the contour in the direction $\text{Re}(s)<0$ and picking up
the poles along the negative real $s$-axis we find the large $\mu$
expansion of $\bra \Tr U\ket^{\text{GC}}$.
Hence the non-trivial information of  instanton corrections is
contained in the pole structure of the {\it spectral trace} $\Tr(\rho^sW)$.
In the classical limit $\hbar\to0$,
the spectral trace $\Tr(\rho^s W)$ can be approximated by the phase space integral
\begin{align}
 Z_0(s)= \int\frac{dxdp}{2\pi\hbar}
\frac{e^{\frac{x+p}{k}}}{(2\cosh\frac{x}{2}\cdot 2\cosh\frac{p}{2})^s}
=\frac{\Ga\Big(\frac{s}{2}+\frac{1}{k}\Big)^2
\Ga\Big(\frac{s}{2}-\frac{1}{k}\Big)^2}{2\pi\hbar\Ga(s)^2}.
\label{eq:Z0}
\end{align}
Note that $Z_0(s)$ has poles at $s=2/k-2\ell$ ($\ell\in\mathbb{N}$)
corresponding to the membrane instanton corrections.
We consider the WKB expansion of the imaginary part of
$\Tr(\rho^s W)$
\begin{align}
\begin{aligned}
 \text{Im} \Tr(\rho^s W)&=Z_0(s)D(s),
\end{aligned}
\end{align}
where $Z_0(s)$ is given by \eqref{eq:Z0} and $D(s)$ represents the $\hbar$-corrections
\begin{align}
 D(s)=\sum_{n=1}^\infty \hbar^{2n-1}D_{2n-1}(s). 
\label{eq:Ds}
\end{align}
Then the imaginary part of 1/6 BPS Wilson loop $\cW_1(\mu,k)$ is 
written as a Mellin-Barnes integral
\begin{align}
 \cW_1(\mu,k)
=\int_{c-\ri\infty}^{c+\ri\infty}\frac{ds}{2\pi i}
\frac{\pi}{\sin\pi s} Z_0(s)D(s) e^{s\mu}.
\end{align}
From the explicit form of the first few terms in the expansion
\eqref{eq:Ds},
we find that $D_{2n-1}(s)$ in \eqref{eq:Ds} has the form
\begin{align}
 D_{2n-1}(s)=\frac{p_{4n-5}(s)}{\prod_{j=0}^{2n-3}(s+j)},\qquad (n\geq2)
\end{align}
where $p_{4n-5}(s)$ is a $(4n-5)^\text{th}$ order polynomial of $s$.
One can easily fix the coefficients in this polynomial $p_{4n-5}(s)$
by matching the value at integer $s$. 
When $s=\ell$ is an integer, the trace
is given by 
\begin{align}
 \Tr(\rho^\ell W)=
\int\frac{dxdp}{2\pi\hbar}
\left(\frac{1}{2\cosh\frac{x}{2}}\star\frac{1}{2\cosh\frac{p}{2}}\right)_\star^\ell\star e^{\frac{x+p}{k}},
\end{align}
where the star-product is defined by
\begin{align}
 f\star g=f\exp\left[\frac{\ri\hbar}{2}\Big(\overleftarrow{\del_x}\overrightarrow{\del_p}
-\overleftarrow{\del_p}\overrightarrow{\del_x}\Big)\right]g.
\end{align}
From this expression,
we can easily compute the $\hbar$-expansion of $\Tr(\rho^\ell W)$.
In this way, we have computed $D_{2n-1}(s)$ up to $n_\text{max}=18$.
The first few terms of $D_{2n-1}(s)$ read
\begin{align}
 \begin{aligned}
  D_1(s)&=\frac{1}{2 k^2 s},\\
D_3(s)&=\frac{k^4 s^3-\left(k^2-8\right) k^2 s^2-8 k^2 s-32}{768 k^6
   s (s+1)},\\
D_5(s)&=\Big(-21 k^8 s^7+6 \left(8-17 k^2\right) k^6 s^6+6 \left(19
   k^4+88 k^2+64\right) k^4 s^4\\
&-64 \left(10 k^4+19
   k^2+88\right) k^2 s^2+1024 \left(k^4+4
   k^2+13\right)+\left(-87 k^8+256 k^6+320 k^4\right) s^5\\
&+32
   \left(3 k^6+6 k^4-52 k^2-80\right) k^2 s^3-128 \left(3
  k^6-9 k^4-32 k^2-40\right) s\Big)\\
  &/
\Big(2949120 k^{10} s (s+1)
   (s+2) (s+3)\Big).
 \end{aligned}
\label{eq:Dn-wkb}
\end{align}

\paragraph{Perturbative part.}
The perturbative part of $\cW_1$
comes from the pole at $s=2/k$.
The residue of the pole at $s=2/k$ is given by 
 \begin{align}
  \begin{aligned}
 &\text{Res}_{s=\frac{2}{k}}\left[\frac{\pi}{\sin\pi s} Z_0(s) D(s)e^{s\mu}\right]\\
=&\frac{e^{\frac{2\mu}{k}}}{\pi k\sin\frac{2\pi}{k}}
\left[D(2/k)\Bigl(\mu-\pi \cot\frac{2\pi}{k}-\psi(2/k)-\ga\Bigr)+D'(2/k)\right],  
  \end{aligned}
 \end{align}
where $\psi(2/k)$ denotes the digamma function
and $D'(s)=\del_s D(s)$. 
From the explicit form of the coefficients $D_{2n-1}(s)$ in \eqref{eq:Dn-wkb}, we find that the WKB expansion of $D(2/k)$ becomes
\begin{align}
 D(2/k)=\frac{\hbar}{4k}-\frac{1}{3!}\left(\frac{\hbar}{4k}\right)^3+\frac{1}{5!}\left(\frac{\hbar}{4k}\right)^5+\cdots,
\end{align}
from which we can easily guess the exact form of $D(2/k)$
\begin{align}
D(2/k)=\sin\left(\frac{\hbar}{4k}\right).
\label{eq:D-sin}
\end{align}
We have checked that the WKB expansion of $D(2/k)$ agrees with \eqref{eq:D-sin}
up to $\mathcal{O}(\hbar^{2n_\text{max}-1})=\mathcal{O}(\hbar^{35})$.
Plugging the relation $\hbar=2\pi k$ into \eqref{eq:D-sin}, we find
\begin{align}
 D(2/k)=1.
\end{align}
This correctly reproduces the coefficient of $\mu$
in the perturbative part \eqref{eq:Wn-pert} for the winding number $n=1$.

\begin{figure}[tb]
\begin{center}
\includegraphics[width=6cm]{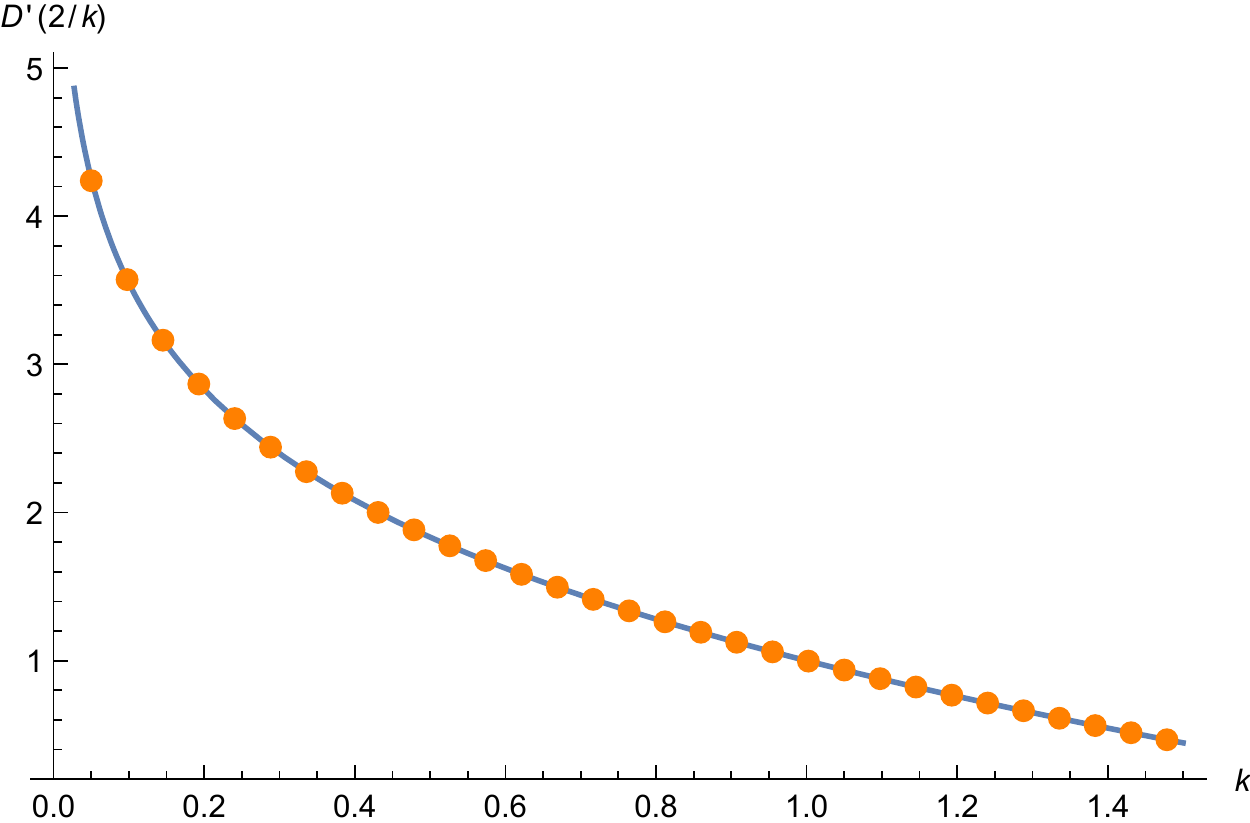}
\end{center}
  \caption{
This is  the plot 
of $D'(2/k)$ against $k$.
The orange dots are the 
numerical values obtained from the Pad\'e approximation,
while
the solid curve is the exact
function $\psi(2/k)+\ga$ in \eqref{eq:m0D}.
}
  \label{fig:m0-D}
\end{figure}

In order to reproduce the $\mathcal{O}(\mu^0)$ 
term in \eqref{eq:Wn-pert}, we need to show the relation
\begin{align}
D'(2/k)=\psi(2/k)+\ga.
\label{eq:m0D}
\end{align}
We do not have an analytic proof of this relation, but we can
check this
using the diagonal Pad\'e approximation 
\begin{align}
 D'(2/k)\approx\sum_{n=1}^{n_\text{max}}D_{2n-1}'(2/k)\hbar^{2n-1}
\approx\hbar\frac{a_0+a_1\hbar^2+\cdots a_{\frac{n_\text{max}}{2}-1}\hbar^{n_\text{max}-2}}
{1+b_1\hbar^2+\cdots b_{\frac{n_\text{max}}{2}-1}\hbar^{n_\text{max}-2}}.
\end{align}
As we can see from Fig.~\ref{fig:m0-D},  the Pad\'e approximation 
of $D'(2/k)$ nicely agrees with the expected 
function \eqref{eq:m0D}.

\begin{figure}[thb]
\begin{center}
\begin{tabular}{cc}
\hspace{-4mm}
\includegraphics[width=6cm]{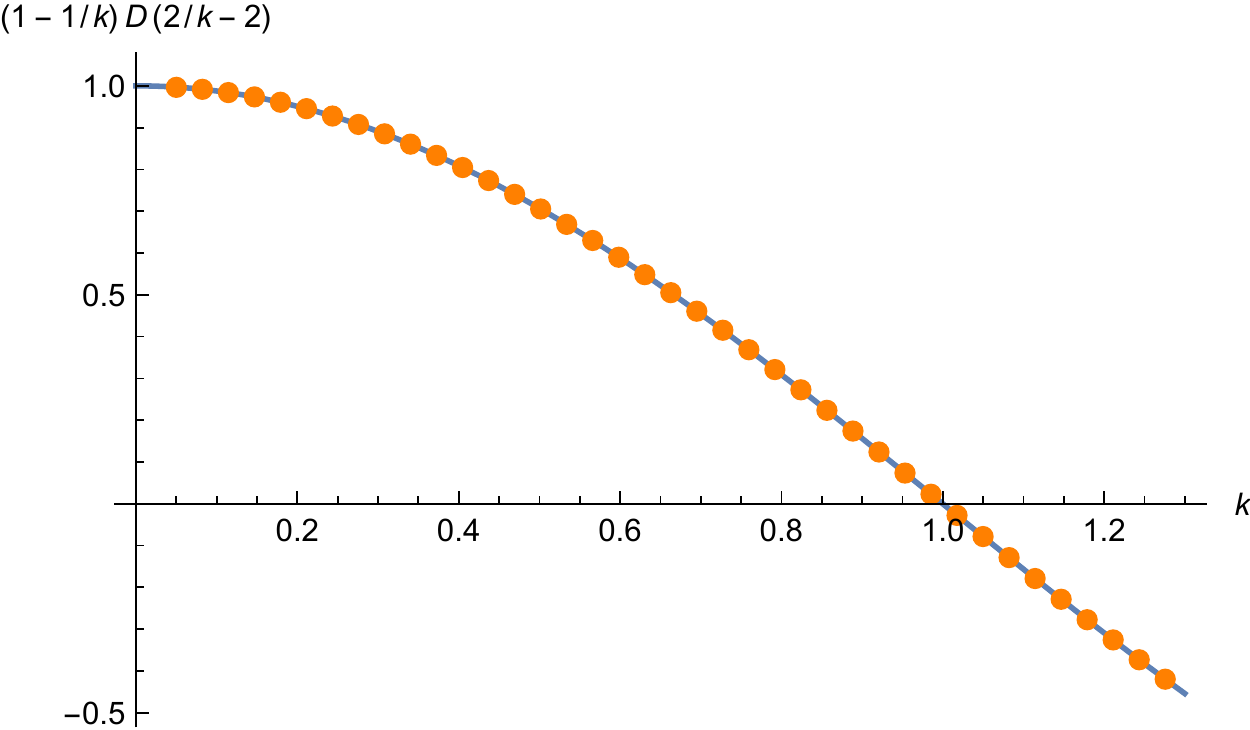}
\hspace{10mm}
&
\includegraphics[width=6cm]{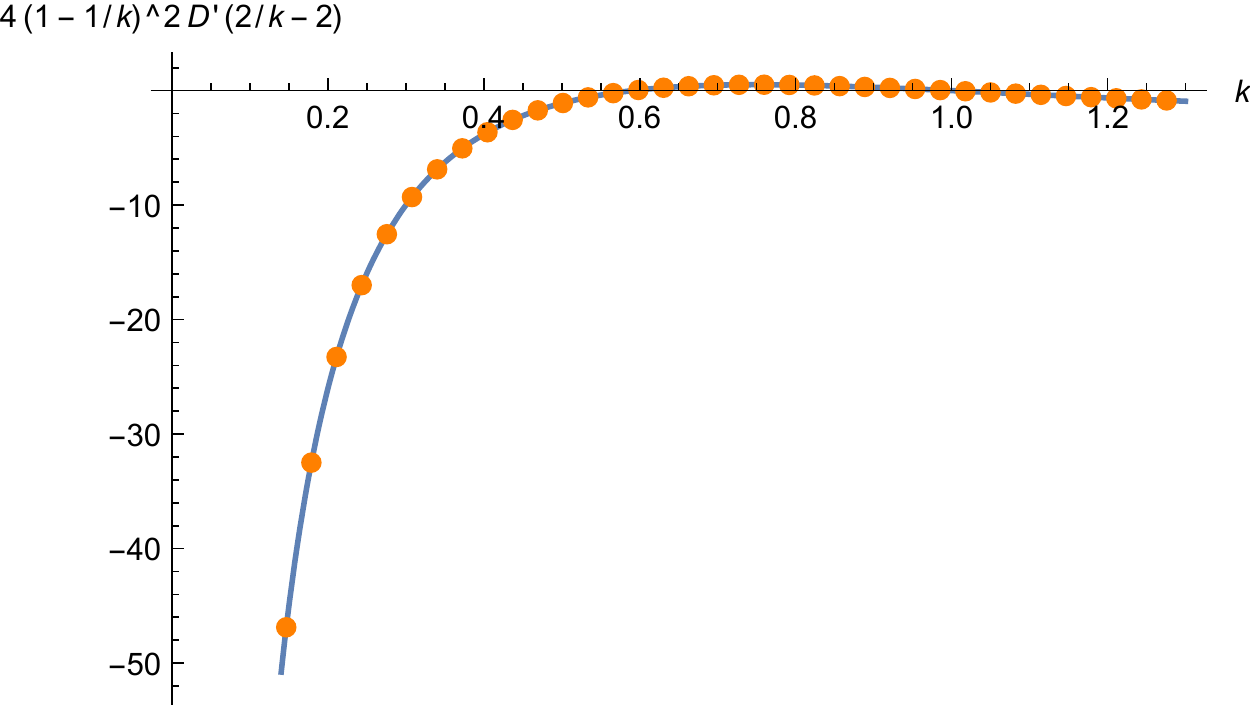}\\
(a) $(1-1/k)D(2/k-2)$
\hspace{10mm}
&
(b) $4(1-1/k)^2D'(2/k-2)$
\end{tabular}
\end{center}
  \caption{
We show the plot 
of (a) $(1-1/k)D(2/k-2)$ and (b) $4(1-1/k)^2D'(2/k-2)$.
The orange dots are the 
numerical values obtained from the Pad\'e approximation,
while
the solid curves represent 
the exact functions in \eqref{eq:D-mem1}.
}
  \label{fig:m1-D}
\end{figure}

\paragraph{Membrane 1-instanton.}
The membrane 1-instanton term comes from the pole at $s=2/k-2$.
In order to reproduce
the result in \eqref{eq:mem-1}, we find that 
$D(2/k-2)$ and $D'(2/k-2)$ should satisfy
\begin{align}
\begin{aligned}
 \Bigl(1-\frac{1}{k}\Bigr)D\left(\frac{2}{k}-2\right)&=\cos\frac{\pi k}{2},\\
 4\Bigl(1-\frac{1}{k}\Bigr)^2D'\left(\frac{2}{k}-2\right)&=
-\Bigl(2+\pi k\cot\frac{\pi k}{2}\Bigr)\cos\frac{\pi k}{2}\\
&+
4\Bigl(1-\frac{1}{k}\Bigr)\Bigl(-1+\ga+2\psi(2/k-2)-\psi(2/k-1)\Bigr)\cos\frac{\pi k}{2}.
\end{aligned}
\label{eq:D-mem1}
\end{align}
As in the case of perturbative part, we can check these relations numerically by the Pad\'e approximation.
From Fig.~\ref{fig:m1-D},
one can see that the Pad\'e approximation of $D(2/k-2)$ and $D'(2/k-2)$ agrees 
with the exact functions \eqref{eq:D-mem1}.

\begin{figure}[thb]
\begin{center}
\begin{tabular}{cc}
\hspace{-4mm}
\includegraphics[width=6cm]{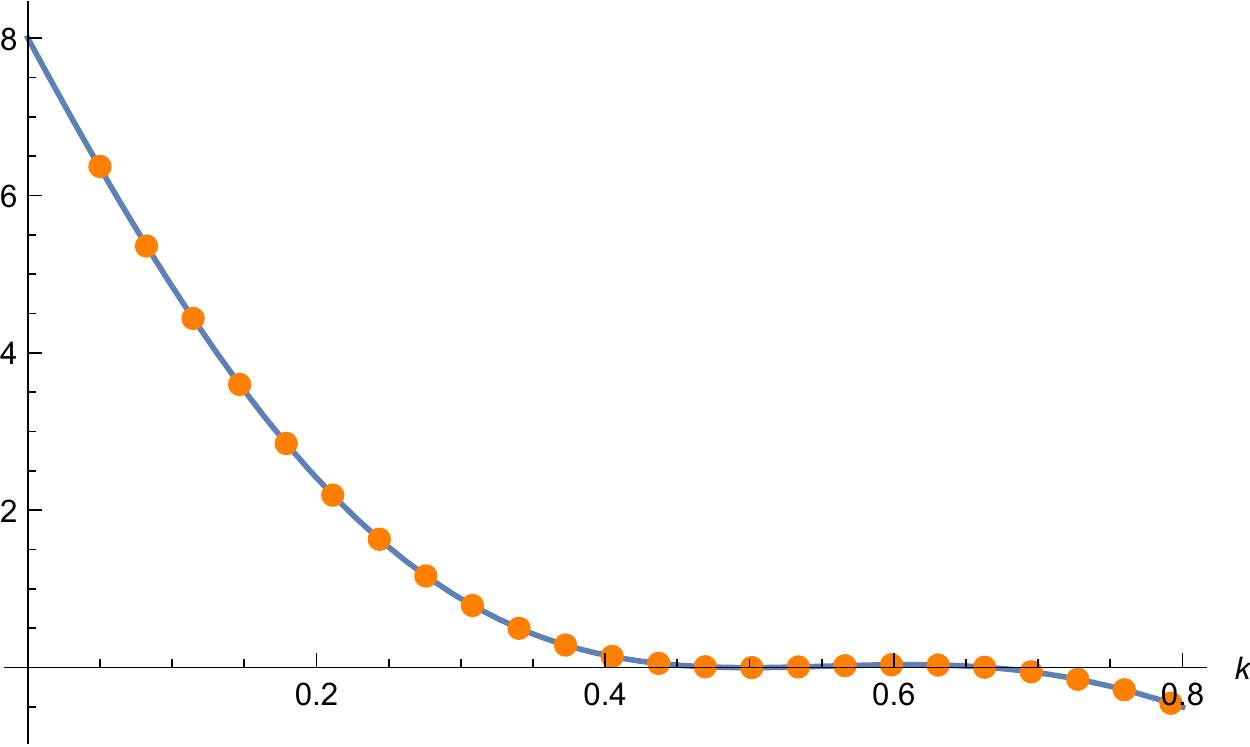}
\hspace{10mm}
&
\includegraphics[width=6cm]{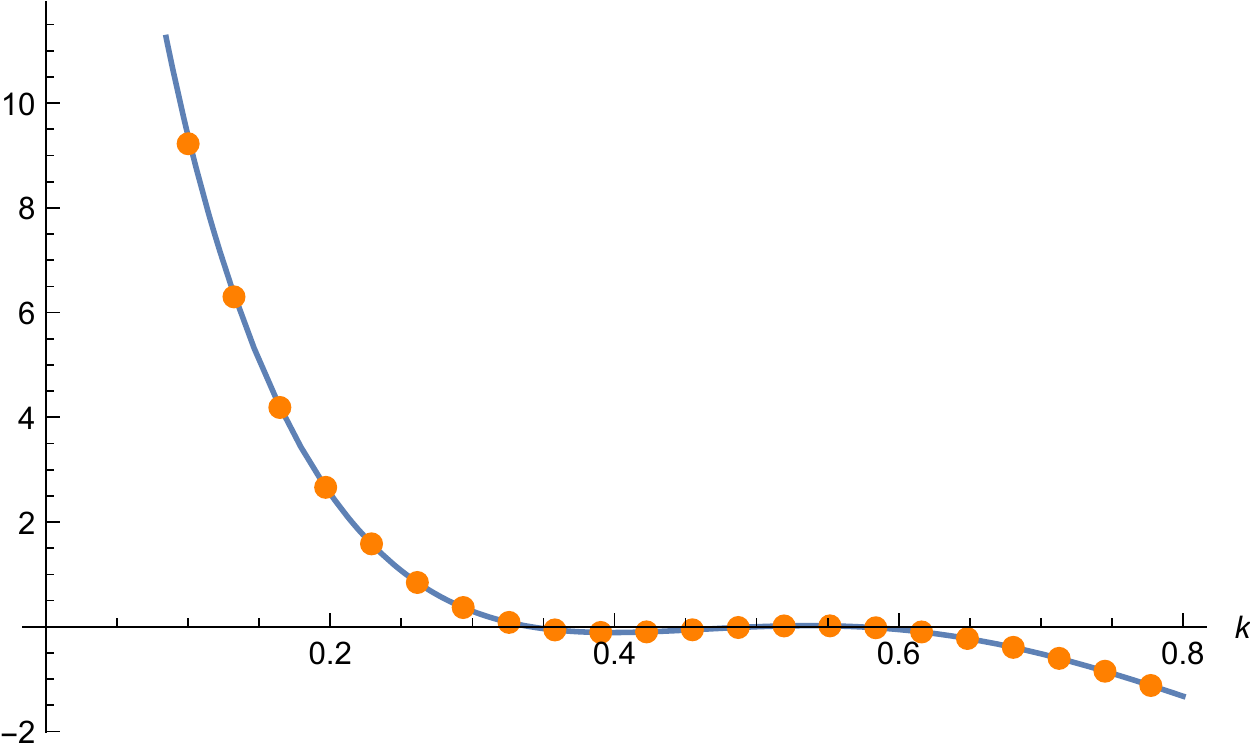}\\
(a) $\frac{k^2\Ga\left(2/k-2\right)^2}{4\Ga\left(2/k-4\right)^2} D\left(2/k-4\right)$
\hspace{10mm}
&
(b) $\frac{k^2\Ga\left(2/k-2\right)^2}{4\Ga\left(2/k-4\right)^2} D'\left(2/k-4\right)$
\end{tabular}
\end{center}
  \caption{
We show the plot 
of (a) $\frac{k^2\Ga\left(2/k-2\right)^2}{4\Ga\left(2/k-4\right)^2} D\left(2/k-4\right)$ and (b) $\frac{k^2\Ga\left(2/k-2\right)^2}{4\Ga\left(2/k-4\right)^2} D'\left(2/k-4\right)$.
The orange dots are the 
numerical values obtained from the Pad\'e approximation,
while
the solid curves represent 
the exact functions in \eqref{eq:D-mem2}.
}
  \label{fig:m2-D}
\end{figure}

\paragraph{Membrane 2-instanton.}
We can repeat the same analysis for the membrane 2-instanton.
In order to reproduce the result in the previous section, we need the following relations
\begin{align}
 \begin{aligned}
 \frac{k^2\Ga\left(\frac{2}{k}-2\right)^2}{4\Ga\left(\frac{2}{k}-4\right)^2} D\left(\frac{2}{k}-4\right)=&2 (2k-1) (4 k-2+(5 k-2) \cos \pi k),\\
\frac{k^2\Ga\left(\frac{2}{k}-2\right)^2}{4\Ga\left(\frac{2}{k}-4\right)^2} D'\left(\frac{2}{k}-4\right)=&-\frac{1}{2} \pi  k^2 \left((4-18 k) \sin \pi  k+k \tan
  \frac{\pi  k}{2}+(25 k-8) \cot
  \frac{\pi  k}{2}\right)\\
&-k (8 k-4+(9 k-4) \cos\pi  k)\\
&+2 (2k-1) (4 k-2+(5 k-2) \cos \pi k)(3-2\ga-4\psi(2/k-4)+2\psi(2/k-2)).
 \end{aligned}
\label{eq:D-mem2}
\end{align}
Again,
the Pad\'e approximation of the WKB expansion of 
$D(2/k-4)$ and $D'(2/k-4)$ correctly reproduces
the expected analytic functions on the right hand side of \eqref{eq:D-mem2} (see Fig.~\ref{fig:m2-D}).

\subsection{Comment on winding Wilson loops}
One can in principle compute the WKB expansion of 
winding Wilson loops $\cW_n(\mu,k)$ with $n\geq2$
in a similar manner as the $n=1$ case studied above.
Then $\cW_n(\mu,k)$ is written as a Mellin-Barnes type integral
\begin{align}
 \cW_n(\mu,k)=\int_{c-\ri\infty}^{c+\ri\infty}\frac{ds}{2\pi i}
\frac{\pi}{\sin\pi s} Z_0^{(n)}(s)D^{(n)}(s)e^{s\mu},
\end{align}
where the classical trace $Z_0^{(n)}(s)$ is given by
\begin{align}
 Z_0^{(n)}(s)=\frac{\Ga\left(\frac{s}{2}+\frac{n}{k}\right)\Ga\left(\frac{s}{2}-\frac{n}{k}\right)}{2\pi\hbar\Ga(s)^2},
\end{align}
and $D^{(n)}(s)$ represents the $\hbar$-corrections.
Let us consider the perturbative part coming from the pole at $s=2n/k$
\begin{align}
\begin{aligned}
 &\text{Res}_{s=\frac{2n}{k}}\left[\frac{\pi}{\sin\pi s} Z_0^{(n)}(s) D^{(n)}(s)e^{s\mu}\right]\\
=&\frac{e^{\frac{2n\mu}{k}}}{\pi k\sin\frac{2\pi n}{k}}
\left[D^{(n)}\Bigl(\frac{2n}{k}\Bigr)\Bigl(\mu-\pi \cot\frac{2\pi n}{k}-\psi\Bigl(\frac{2n}{k}\Bigr)-\ga\Bigr)+D^{(n)'}\Bigl(\frac{2n}{k}\Bigr)\right]. 
\end{aligned}
\end{align}
One can easily show that $D^{(n)}(2n/k)=1$ as in the case of $n=1$, 
and hence the coefficient of $\mu$ in the perturbative part \eqref{eq:Wn-pert}
is correctly reproduced from the WKB expansion.
In order to reproduce the $\mathcal{O}(\mu^0)$ term in \eqref{eq:Wn-pert},
we have to show that
\begin{align}
 D^{(n)'}\Bigl(\frac{2n}{k}\Bigr)=\psi\Bigl(\frac{2n}{k}\Bigr)+\ga
-\pi\sum_{j=1}^{n-1}\cot\frac{2\pi j}{k}.
\label{eq:dense}
\end{align}
However, it is not so straightforward to reproduce
this behavior from the Pad\'e approximation of WKB expansion,
since  the function on the right hand side of \eqref{eq:dense} has 
a dense set of poles at rational values of $k$, accumulating around $k=0$.
Note that this problem does not occur for $n=1$.
At present we could not succeed in  reproducing the perturbative part of
winding Wilson loop from the WKB expansion for $n\geq2$.
It would be interesting to derive the perturbative part \eqref{eq:Wn-pert}
of winding Wilson loop analytically.

\section{Winding Wilson Loops}\label{sec:wind}
In this section, we will consider the
instanton corrections of the
imaginary part of 1/6 BPS winding Wilson loops 
$\cW_n(\mu,k)$ with the winding number $n\geq2$.
We can study the the instanton corrections of $\cW_n(\mu,k)$ for
 $n\geq2$ 
by the same method of numerical fitting we used for the fundamental representation 
in section \ref{sec:fund}.
\subsection{Winding number $n=2$}
Let us first consider the Wilson loop with winding number $n=2$.
After rewriting $\cW_2(\mu,k)$ in terms of  the effective chemical potential 
$\mu_\text{eff}$, we find that
the factorized structure \eqref{eq:fact-fund}
found in the fundamental representation
also appears in the $n=2$ case
\begin{align}
 \cW_2=\frac{e^{\frac{4\mu_\text{eff}}{k}}}{k\pi\sin\frac{4\pi}{k}}\frac{\del\mu_\text{eff}}{\del\mu}
f_w\left[\mu_\text{eff}-\pi\cot\frac{2\pi}{k}V_w -\pi\cot\frac{4\pi}{k}f_w^{-1}-\pi k V_m\right],
\label{eq:W2-inst}
\end{align}
where
\begin{align}
 \begin{aligned}
  f_w&=1-4\cos^2\frac{2\pi}{k}Q_w+\frac{2\sin\frac{6\pi}{k}}{\sin\frac{2\pi}{k}}Q_w^2
-16\cos^2\frac{2\pi}{k}Q_w^3\\
&\hskip10mm+\Big(33+40\cos\frac{4\pi}{k}+6\cos\frac{8\pi}{k}\Big)Q_w^4
+\mathcal{O}(Q_w^5),\\
V_w&=1+2\cos\frac{4\pi}{k}Q_w+4\cos^2\frac{4\pi}{k}Q_w^2+8\cos^2\frac{4\pi}{k}
\Bigl(\cos\frac{4\pi}{k}+\frac{2\sin\frac{2\pi}{k}}{\sin\frac{6\pi}{k}}\Bigr)Q_w^3\\
&+\Bigl(9-4 \cos \frac{4 \pi }{k}+6 \cos \frac{8
   \pi }{k}+2 \cos \frac{16 \pi }{k}+\frac{2}{\cos^2\frac{2 \pi }{k}}-\frac{1}{\cos\frac{4 \pi }{k}}
   \Bigr)Q_w^4+\mathcal{O}(Q_w^5),\\
V_m&=\cot\frac{\pi k}{2}\cos\frac{\pi k}{2}Q_m+(4+5\cos\pi k)\cot\pi k Q_m^2+\mathcal{O}(Q_m^3).
 \end{aligned}
 \label{eq:n=2Vm}
\end{align}
Interestingly, the membrane 
instanton correction $V_m$ in \eqref{eq:n=2Vm}
is exactly the same as that of the fundamental Wilson loop.
From this observation, we conjecture that $V_m$ for $n=2$ is again
given by the NS free energy \eqref{eq:Vm-FNS}.
One can show that poles in the convergence region $k>4$
are canceled between worldsheet instantons and membrane instantons,
and the remaining finite
part reproduces the result in \eqref{eq:w2-int-k}.
This implies that the worldsheet instanton corrections are essentially given by
the S-dual of NS free energy \eqref{eq:Om-np}, up to regular terms.

\subsection{Winding number $n\geq3$}
Let us consider the Wilson loops with winding number $n\geq3$.
From the convergence condition $k>2n$,
as the winding number increases 
the membrane instanton correction
$e^{-2\mu}$ becomes highly suppressed relative to the
worldsheet instanton correction
\begin{align}
 e^{-\frac{4\mu}{k}}\gg e^{-2\mu}.
\end{align}
Because of this, it becomes harder and harder to study membrane instanton
corrections numerically as the winding number
increases. We were unable to find the
analytic expressions of the membrane instanton coefficients for $n\geq3$.

On the other hand, we can determine 
the worldsheet instanton corrections of $\cW_n(\mu,k)$
from numerical fitting.
For $n=3$, the worldsheet instanton corrections are given by
\begin{align}
 \begin{aligned}
  \cW_3^{\text{WS}}&=\frac{e^{\frac{6\mu}{k}}}{k\pi\sin\frac{6\pi}{k}}\Biggl[f_w\Big(\mu-
\pi\cot\frac{2\pi}{k}V_w\Big)-\pi\cot\frac{4\pi}{k}-\pi\cot\frac{6\pi}{k}\Biggr],\\
f_w&=1-\frac{\sin^2\frac{6\pi}{k}}{\sin^2\frac{2\pi}{k}}Q_w+3\frac{\sin^2\frac{6\pi}{k}}{\sin^2\frac{2\pi}{k}}Q_w^2+\left(2-9\frac{\sin^2\frac{6\pi}{k}}{\sin^2\frac{2\pi}{k}}\right)Q_w^3
+\mathcal{O}(Q_w^4),\\
V_w&=1+\frac{\cos\frac{4\pi}{k}\sin\frac{6\pi}{k}}{\cos^2\frac{2\pi}{k}\sin\frac{2\pi}{k}}Q_w
+\frac{\sin\frac{6\pi}{k}}{\cos^2\frac{2\pi}{k}\sin\frac{2\pi}{k}}\left(\frac{3}{2}+
\cos\frac{4\pi}{k}+\cos\frac{8\pi}{k}+\cos\frac{12\pi}{k}\right)Q_w^2
+\mathcal{O}(Q_w^3),
 \end{aligned}
\end{align}
and for $n=4$, worldsheet instanton corrections are given by
\begin{align}
 \begin{aligned}
  \cW_4^{\text{WS}}&=\frac{e^{\frac{8\mu}{k}}}{k\pi\sin\frac{8\pi}{k}}\Biggl[f_w\Big(\mu-
\pi\cot\frac{2\pi}{k}V_w\Big)-\pi\cot\frac{4\pi}{k}-\pi\cot\frac{6\pi}{k}
-\pi\cot\frac{8\pi}{k}\Biggr],\\
f_w&=1-\frac{\sin^2\frac{8\pi}{k}}{\sin^2\frac{2\pi}{k}}Q_w
+4\cos^2\frac{4\pi}{k}\Bigl(7+8\cos\frac{4\pi}{k}+2\cos\frac{8\pi}{k}\Bigr)Q_w^2
+\mathcal{O}(Q_w^3),\\
V_w&=1+4\frac{\cos\frac{4\pi}{k}\sin\frac{2\pi}{k}}{\sin\frac{6\pi}{k}}\Bigl(\cos\frac{4\pi}{k}
+\cos\frac{8\pi}{k}-\cos\frac{12\pi}{k}\Bigr)Q_w
+\mathcal{O}(Q_w^2).
 \end{aligned}
\end{align}

We find that the worldsheet 1-instanton correction 
in the coefficient of $\mu$ can be written in a closed form
for general $n\geq2$
\begin{align}
 f_w=1-\frac{\sin^2\frac{2n\pi}{k}}{\sin^2\frac{2\pi}{k}}Q_w+\mathcal{O}(Q_w^2).
\label{eq:fw-n}
\end{align}

It would be very interesting to understand the structure
of instanton correction of $\cW_n(\mu,k)$ and see if it is related to
the quantum volume for the general winding number $n$.
\section{Wilson loops with two boundaries}\label{sec:two-b}
As pointed out in \cite{KMSS}, 
1/6 BPS Wilson loops in higher rank representation
involves multi-particle interaction in the 
Fermi gas picture, hence it is not so easy
to study the 1/6 BPS Wilson loops in general representations.
Here we initiate the study of Wilson loop with two boundaries
$\bra (\Tr U)^2\ket$, which is related to
the second rank symmetric and anti-symmetric representations by
\begin{align}
 \begin{aligned}
  W_{\tableau{2}}&=W_{S_2}=\hf \bra (\Tr U)^2\ket+\hf \bra \Tr U^2\ket,\\
W_{\tableau{1 1}}&=W_{A_2}=\hf \bra (\Tr U)^2\ket-\hf \bra \Tr U^2\ket.
 \end{aligned}
\end{align}
Note that $\bra \Tr U^2\ket$ is the 1/6 BPS Wilson loop with winding number $n=2$ 
which we have studied
in the previous section.

From the general formula \eqref{eq:det-f},
the generating function of
1/6 BPS Wilson loops in the anti-symmetric representations is given by
\begin{align}
 \sum_{n}t^n W_{A_n}=\frac{\det(1+\ka\rho(1+tW))}{\det(1+\ka\rho)},
\end{align}
while the  generating function of
1/6 BPS Wilson loops in the symmetric representations is given by
\begin{align}
 \sum_{n}t^n W_{S_n}=\frac{\det(1+\ka\rho(1-tW)^{-1})}{\det(1+\ka\rho)}.
\end{align}
Picking up the coefficient of $t^2$ in the above expressions,
we find that $W_{A_2}$ and $W_{S_2}$ are written in terms of $R$
defined in \eqref{eq:R-def}:
 \begin{align}
  \begin{aligned}
   W_{A_2}&=\hf (\Tr RW)^2-\hf \Tr (RW)^2,\\
 W_{S_2}&=\Tr RW^2+ \hf(\Tr RW)^2-\hf \Tr (RW)^2.   
  \end{aligned}
  \label{eq:WA2S2}
 \end{align}
 Note that $\Tr RW$ and
 $\Tr RW^2$ appearing in \eqref{eq:WA2S2}
 are the grand canonical VEV of 1/6 BPS winding Wilson loops
 with the winding numbers $n=1,2$, respectively
\begin{align}
 \Tr RW=\bra \Tr U\ket^\text{GC},\qquad
 \Tr RW^2=\bra \Tr U^2\ket^\text{GC}.
\end{align}
The last term $\Tr (RW)^2$ in \eqref{eq:WA2S2}
is the new contribution which we should compute.
It turns out that
the grand canonical VEV of $(\Tr U)^2$, which we denote
by $W_{(1,1)}$, has
a simple expansion in the large $\mu$ limit
\begin{align}
 W_{(1,1)}=\frac{1}{2}\bra (\Tr U)^2\ket^{\text{GC}}=\hf(W_{S_2}+W_{A_2}).
\end{align}
From \eqref{eq:WA2S2}, $W_{(1,1)}$ is written as
\begin{align}
 W_{(1,1)}=\hf\Big[\Tr RW^2+(\Tr RW)^2- \Tr (RW)^2\Big].
 \label{eq:W11}
\end{align}
\subsection{Computation of $\Tr (RW)^2$}\label{sec:comp-two}
To study the 1/6 BPS Wilson loop with two boundaries
\eqref{eq:W11}, we have to compute $\Tr (RW)^2$.
The first two terms in \eqref{eq:W11}
have been already obtained in the previous sections. 
Now let us expand $\Tr (RW)^2$ in $\ka$ as
\begin{align}
 \Tr (RW)^2=\sum_{\ell=2}^\infty (-1)^\ell\ka^\ell t_\ell,
\end{align}
where $t_\ell$ is given by
\begin{align}
 t_\ell=\sum_{j=1}^{\ell-1}\Tr(\rho^j W\rho^{\ell-j}W).
\label{eq:def-tl}
\end{align}
To compute $t_\ell$, we rewrite the trace in \eqref{eq:def-tl} as
\begin{align}
 \Tr(\rho^j W\rho^{\ell-j}W)=\Tr(\rho^{\ell}W^2)+
 \hf \Tr([\rho^j,W][\rho^{\ell-j},W]).
 \label{eq:rho-comm}
\end{align}
Note that the first term $\Tr(\rho^{\ell}W^2)$ in \eqref{eq:rho-comm}
is the trace appearing in the winding Wilson loop
with winding number $n=2$.
The second term in \eqref{eq:rho-comm} can be further simplified
by noticing that
\begin{align}
[\rho,W]&=\frac{1}{2\cosh\frac{x}{2}}\left\{\frac{1}{2\cosh\frac{p}{2}},W\right\}-\left\{\frac{1}{2\cosh\frac{x}{2}},W\right\}\frac{1}{2\cosh\frac{p}{2}},
\label{eq:com-rho}
\end{align}
and using the fact that the anti-commutators in \eqref{eq:com-rho} have
a simple matrix element \cite{HMMO-Wilson}
\begin{align}
 \begin{aligned}
  \bra x_1|\left\{\frac{1}{2\cosh\frac{p}{2}},W\right\}|x_2\ket&=\frac{1}{\hbar}e^{\frac{x_1+x_2}{2k}},\\
\bra p_1|\left\{\frac{1}{2\cosh\frac{x}{2}},W\right\}|p_2\ket&=\frac{1}{\hbar}e^{\frac{p_1+p_2}{2k}}.
 \end{aligned}
\end{align}
Then one can show that $t_\ell$ is written as
\begin{align}
\begin{aligned}
 t_\ell
 =(\ell-1)\Tr(\rho^\ell W^2)+\sum_{j=1}^{\ell-1}
 \sum_{a=0}^{j-1}\sum_{b=0}^{\ell-j-1}
(I_{1/2}^{(a+b)}I_{1/2}^{(\ell-2-a-b)}-\bar{I}_{1/6}^{(a+b+1)}I_{1/6}^{(\ell-2-a-b)}), 
\end{aligned}
\end{align}
where
\begin{align}
 \begin{aligned}
  I_{1/2}^{(\ell)}=&\int \frac{dxdy}{\hbar} \,e^{\frac{x}{2k}}\bra x|\rho^\ell|y\ket 
\frac{e^{\frac{y}{2k}}}{2\cosh\frac{y}{2}},\\
I_{1/6}^{(\ell)}=&\int\frac{dxdp}{\hbar} \, e^{\frac{x}{2k}}\bra x|\rho^\ell|p\ket  e^{\frac{p}{2k}}
=\int dx\, \frac{e^{\frac{x}{2k}}}{\rt{k}}\bra x|\rho^\ell|\pi\ri\ket.
 \end{aligned}
 \label{eq:I1/6}
\end{align}
The integral $I_{1/2}^{(\ell)}$ has  appeared
in the computation of 1/2 BPS Wilson loops in a hook representation
\cite{HMMO-Wilson}.
As shown in \cite{HMMO-Wilson},
$I_{1/2}^{(\ell)}$ can be computed systematically by
 constructing a sequence of functions.
 Using the algorithm in section \ref{sec:num}, we can numerically evaluate $I_{1/2}^{(\ell)}$  with high precision.
Similarly, $I_{1/6}^{(\ell)}$ in \eqref{eq:I1/6} can  be computed
 by constructing a sequence of functions
 $\psi_\ell(x)$
 \begin{align}
 \begin{aligned}
  I_{1/6}^{(0)}&=\frac{1}{\rt{k}}e^{\frac{\pi \ri}{2k}},\\
   I_{1/6}^{(\ell)}&=\int dx e^{\frac{x}{2k}} \psi_\ell(x),\quad(\ell\geq1),
 \end{aligned}
 \end{align}
 where $\psi_\ell(x)$ is defined recursively
  \begin{align}
   \begin{aligned}
\psi_\ell(x)=\frac{1}{2\cosh\frac{x}{2}}\int\frac{dy}{2\pi k}\frac{1}{2\cosh\frac{x-y}{2k}}\psi_{\ell-1}(y),\quad
\psi_1(x)=\frac{1}{\rt{k}}\frac{1}{2\cosh\frac{x}{2}}\frac{1}{2\cosh\frac{x-\pi\ri}{2k}}.      
   \end{aligned}
  \end{align}
Again, we can compute $I_{1/6}^{(\ell)}$ numerically
using the algorithm in section \ref{sec:num}.

For some values of integer $k$,
we can compute the trace $t_\ell$
exactly by closing the contour and picking up the residue of poles.
For $k=6$ we find
\begin{align}
 \begin{aligned}
  t_2&=\frac{e^{\frac{2\pi\ri}{6}}}{648}\Biggl(5-\frac{6\rt{3}}{\pi}\Biggr),\\
 t_3&=\frac{-972+216 \ri \sqrt{3}+180 \ri \pi -864 \sqrt{3} \pi +585
   \pi ^2-46 \ri \sqrt{3} \pi ^2}{93312 \pi ^2}.
 \end{aligned}
\end{align}
For $k=8$ we find
\begin{align}
 \begin{aligned}
 t_2&=\frac{e^{\frac{2 \pi \ri}{8}} \left(2
   \sqrt{2}+\left(\sqrt{2}-2\right) \pi \right)}{256 \pi },\\
t_3&=\frac{-4+(16+4 \ri) \pi +(32-16 \ri) \sqrt{2} \pi -(13-12 \ri) \pi
   ^2-(4+4 \ri) \sqrt{2} \pi ^2}{8192 \pi ^2}.
 \end{aligned}
\end{align}
For $k=12$ we find
\begin{align}
 t_2&=e^{\frac{2\pi\ri}{12}}\frac{18-27 \pi +13 \sqrt{3} \pi }{2592 \pi },\\
t_3&= \frac{-54+\left((486+54 \ri)+(120-216 \ri) \sqrt{3}\right) \pi
   +\left((-209-279 \ri)+222 \ri \sqrt{3}\right) \pi ^2}{373248
   \pi ^2}.
\end{align}
We have checked that the numerical computation
using the method in section \ref{sec:num}
correctly reproduces the exact values of $t_\ell$
for $k=6,8,12$ with high precision.
We should stress that we can compute $t_\ell$
numerically at arbitrary $k$ 
in the convergence region $k>4$.

\subsection{Imaginary part of $W_{(1,1)}$}
Let us first consider the
imaginary part of $W_{(1,1)}$, which we denote
by $W_{(1,1)}^{\text{Im}}$.
From the numerical fitting, we find that the perturbative part of $W_{(1,1)}^{\text{Im}}$ is given by
\begin{align}
 W_{(1,1)}^{\text{Im,pert}}=\frac{e^{\frac{4\mu}{k}}}{\pi k(2\sin\frac{2\pi}{k})^2}\Bigl(\mu
-2\pi \cot\frac{4\pi}{k}\Bigr).
\end{align}
We conjecture that the instanton corrections
have the following structure
\begin{align}
\begin{aligned}
 W_{(1,1)}^{\text{Im}}=&\frac{e^{\frac{4\mu_\text{eff}}{k}}}{\pi k(2\sin\frac{2\pi}{k})^2}
f_w\frac{\del\mu_\text{eff}}{\del\mu}\Biggl(\mu_\text{eff}
-2\pi\cot\frac{4\pi}{k}\frac{V_4}{f_w}
+2\pi\cot\frac{2\pi}{k}V_2-\pi kV_m\Biggr) ,
\end{aligned}
\label{eq:W11-im}
\end{align}
where the membrane instanton correction $V_m$ is given by the NS free energy \eqref{eq:Vm-FNS}, and
 the worldsheet instanton corrections are given by
\begin{align}
 \begin{aligned}
  f_w&=1+4\sin^2\frac{2\pi}{k}Q_w-2\Bigl(1+4\sin^2\frac{2\pi}{k}\Bigr)Q_w^2
+16\sin^2\frac{2\pi}{k}Q_w^3+\Bigl(15-96\sin^2\frac{2\pi}{k}+48\sin^4\frac{2\pi}{k}\Bigr)Q_w^4,\\
V_4&=1-Q_w^2+4Q_w^3,\\
V_2&=Q_w-4\sin^2\frac{2\pi}{k}Q_w^2+\frac{2\sin\frac{2\pi}{k}}{\sin\frac{6\pi}{k}}\Bigl(-3+4\cos\frac{4\pi}{k}-6\cos\frac{8\pi}{k}+\cos\frac{12\pi}{k}\Bigr) Q_w^3.
 \end{aligned}
\end{align}
We have checked that the membrane 1-instanton and $(1,1)$-boundstate
are correctly reproduced from \eqref{eq:W11-im}.
Also, one can check that the pole cancellation between worldsheet instantons and membrane instantons works for $k=6,8,12$, and the remaining finite terms
correctly reproduce the result in Appendix \ref{app:Im}.

\subsection{Real part of $W_{(1,1)}$}
Next consider the real part of $W_{(1,1)}$,
which we denote by $W_{(1,1)}^\text{Re}$.
Again, from the numerical fitting,
we find
that the perturbative part of $W_{(1,1)}^\text{Re}$ is given by
\begin{align}
W_{(1,1)}^{\text{Re},\text{pert}}=
 \frac{ e^{\frac{4\mu}{k}}}{(2\sin\frac{2\pi}{k})^2}
\left[-\frac{2}{\pi^2k^2}\Bigl(\mu-2\pi\cot\frac{4\pi}{k}\Bigr)^2 +\frac{1}{8}-\frac{\tan\frac{2\pi}{k}}{\pi k}\right], 
\end{align}
and the worldsheet instanton corrections are given by
\begin{align}
\begin{aligned}
 e^{-\frac{4\mu_{\text{eff}}}{k}} W_{(1,1)}^{\text{Re},\text{WS}}&=\frac{ Q_w}{\sin^2\frac{2\pi}{k}}
\left[-\frac{2\mu_{\text{eff}}^2}{\pi^2k^2}\Bigl(1+\sin^2\frac{2\pi}{k}\Bigr)
+\frac{2\mu_{\text{eff}}}{\pi k^2}\cot\frac{2\pi}{k}-\frac{2}{k^2}
+\frac{\sin\frac{4\pi}{k}}{2\pi k}\right] \\
&+\frac{ Q_w^2}{\sin^2\frac{2\pi}{k}}\Biggl[
\frac{\mu_{\text{eff}}^2}{\pi^2k^2}\Bigl(-1+4\sin^2\frac{2\pi}{k}\Bigr)+
\frac{2\mu_{\text{eff}}}{\pi k^2}\Bigl(-4\cot\frac{2\pi}{k}+3\cot\frac{4\pi}{k}\Bigr)
\\
&\hskip20mm+\frac{2}{k^2}+\frac{4}{k^2\sin^2\frac{2\pi}{k}}+\frac{3}{16}+\frac{\cos\frac{4\pi}{k}}{8}-\frac{\sin\frac{6\pi}{k}}{2\pi k\cos\frac{2\pi}{k}}\Biggr]\\
&+\frac{ Q_w^3}{\sin^2\frac{2\pi}{k}}\Biggl[-\frac{32\mu_{\text{eff}}^2}{\pi^2 k^2}\sin^2\frac{2\pi}{k}
+\frac{\mu_{\text{eff}}}{\pi k^2}\frac{13 + 13 \cos\frac{4\pi}{k} + 5 \cos\frac{8\pi}{k}- 3 \cos\frac{12\pi}{k}}{\cos\frac{2\pi}{k}\sin\frac{6\pi}{k}}\\
&\hskip20mm+\frac{2\sin\frac{4\pi}{k}}{\pi k}+\mathcal{O}(k^{-2})+\mathcal{O}(k^{0})\Biggr].
\end{aligned}
\label{eq:W11re-WS}
\end{align}
We find that the membrane 1-instanton has the form
\begin{align}
 W_{(1,1)}^{\text{Re,M2-1}}=\frac{e^{\frac{4\mu_{\text{eff}}}{k}}}{\sin^2\frac{2\pi}{k}}\left[
-\frac{4\cos\frac{\pi k}{2}}{\pi^2 k^2}\mu_{\text{eff}}^2+b_1(k)\mu_{\text{eff}}+c_1(k)\right]e^{-2\mu_{\text{eff}}}.
 \label{eq:W11re-mem}
\end{align}
In the limit $k\to6$, we find that the membrane 1-instanton \eqref{eq:W11re-mem}
and the worldsheet 3-instanton in \eqref{eq:W11re-WS}
reproduce the coefficient of $\mu_\text{eff}^2$ term for $k=6$ in \eqref{eq:Wre-k68}
\begin{align}
 \lim_{k\to6}e^{-\frac{4\mu_\text{eff}}{k}}\Bigl[W_{(1,1)}^{\text{Re},\text{M2-1}}+W_{(1,1)}^{\text{Re},\text{WS-3}}\Bigr]=
\left[-\frac{32\mu_{\text{eff}}^2}{52\pi^2}+\mathcal{O}(\mu_{\text{eff}})\right]
e^{-2\mu_{\text{eff}}}.
\label{eq:W11-cancel}
\end{align}
The $\mathcal{O}(\mu_\text{eff})$ term in the worldsheet 3-instanton has a pole at $k=6$
which should be canceled by the membrane 1-instanton.
From this pole cancellation condition, we conjecture
that the membrane 1-instanton coefficient of $\mu_\text{eff}$ in \eqref{eq:W11re-mem}
has the structure
\begin{align}
 b_1(k)=\frac{\cos\frac{\pi k}{2}\cot\frac{\pi k}{2}}{\pi k}+(\text{regular}).
\end{align}
We leave the further study of membrane instanton corrections
of $W_{(1,1)}^\text{Re}$ as an interesting future problem.

\subsection{Connected part of $W_{(1,1)}$}
We expect that the connected part of $\bra (\Tr U)^2\ket$
has a simpler behavior
\begin{align}
 \hf \bra (\Tr U)^2\ket_\text{conn}=\hf \bra (\Tr U)^2\ket
 -\hf \bra \Tr U\ket^2.
\end{align}
Let us consider the perturbative part
of $\bra \Tr U\ket$ and
$\bra (\Tr U)^2\ket$
\begin{align}
 \begin{aligned}
  \bra \Tr U\ket^\text{pert}&=\frac{e^\frac{2\mu}{k}}{2\sin\frac{2\pi}{k}}
  \left[\hf+\frac{2\ri}{\pi k}\Bigl(\mu-\pi\cot\frac{2\pi}{k}\Bigr)\right],\\
  \hf \bra (\Tr U)^2\ket^\text{pert}&= \frac{e^\frac{4\mu}{k}}
  {(2\sin\frac{2\pi}{k})^2}\left[-\frac{2}{\pi^2 k^2}
  \Bigl(\mu-2\pi\cot\frac{4\pi}{k}\Bigr)^2+\frac{1}{8}-\frac{\tan\frac{2\pi}{k}}{\pi k}+\frac{\ri}{\pi k}\Bigl(\mu-2\pi \cot\frac{4\pi}{k}\Bigr)\right].
 \end{aligned}
\end{align}
From this expression, we  find that the $\mu^2$ term is canceled in the connected part 
of $\bra (\Tr U)^2\ket$
\begin{align}
 \hf \bra (\Tr U)^2\ket_\text{conn}^\text{pert}
 =\frac{e^{\frac{4\mu}{k}}}{2\pi k\sin\frac{4\pi}{k}}\left[
 -\frac{2}{k}\Bigl(2\mu-2\pi \cot\frac{4\pi}{k}-\pi \cot\frac{2\pi}{k}\Bigr)
 -1+\pi\ri\right].
\end{align}
However, this cancellation of $\mu^2$ term does not hold for the worldsheet instanton corrections.
It would be interesting to study the 
structure of the connected part $\bra (\Tr U)^2\ket_\text{conn}$ 
further.

\section{Conclusions}\label{sec:conclude}
In this paper we have studied the instanton corrections to the 
VEV of 1/6 BPS Wilson loops in ABJM theory from Fermi gas approach.
For the fundamental representation, we find that the grand canonical VEV 
of the imaginary part of Wilson loop has a remarkably simple form 
\eqref{eq:W1-final}, and it 
is closely related to the quantum volume
given by the NS free energy
on local $\mathbb{P}^1\times \mathbb{P}^1$. 
The poles at rational value of $k$
in the  membrane instanton corrections are canceled
by the worldsheet instanton corrections and the singular part of worldsheet instantons is
essentially given by the ``S-dual'' of membrane instantons.
This is reminiscent of the exact quantization condition of the spectrum. 
It would be very interesting to understand the physical meaning of this relation
between the quantum volume and the 1/6 BPS Wilson loops.

It is curious that the
NS free energy is determined by the BPS invariant of {\it closed} topological string, while we naively expect that the Wilson loops 
are related to some {\it open} string amplitudes.
This type of {\it open-closed duality} was observed
for the 1/2 BPS Wilson loops in ABJ(M) theory \cite{HO-1/2}.
We speculate that the relation \eqref{eq:W1-final}
between 1/6 BPS Wilson loops and the quantum volume is
a ``1/6 BPS version'' of the open-closed duality.
It would be interesting to study 1/6 BPS Wilson loops in ABJ theory along the lines of \cite{Hirano:2014bia,Matsumoto:2013nya,HO-1/2},
and see if the relation to quantum volume also appears in Wilson loops in ABJ theory.

We have also studied 1/6 BPS winding Wilson loops and initiated
the study of 1/6 BPS Wilson loops with two boundaries.
For both of the imaginary part
of winding Wilson loop $\cW_n$ and
the imaginary part of Wilson loop with two boundaries $W_{(1,1)}^\text{Im}$, we find that the membrane instanton correction
$V_m$ is again given by the NS free energy \eqref{eq:Vm-FNS}.
We also find that the perturbative part of
winding Wilson loop is different from the expression in
\cite{KMSS}. It would be interesting to derive our
result of $\cW_n^\text{pert}$ in \eqref{eq:Wn-pert} analytically.

Our method in section \ref{sec:comp-two} can in principle be generalized to arbitrary number of
boundaries, but the computation becomes cumbersome as the number of boundaries increases. It would be important to develop a more efficient
method to compute the multi-boundary Wilson loops both numerically and analytically.
From our result of $W_{(1,1)}$, it is natural to expect
that as the number of boundaries increases,
the order of polynomial of $\mu_\text{eff}$
in the grand canonical VEV also increases.
It would be interesting to find the general structure
of the grand canonical VEV of 1/6 BPS Wilson loops with
$h$ boundaries.

\acknowledgments
I would like to thank Yasuyuki Hatsuda for
a collaboration during the initial stage of this work.
I would also like to thank Marcos Mari\~no for correspondence.
This work was supported in part by JSPS KAKENHI Grant Number
16K05316, and JSPS Japan-Hungary and Japan-Russia bilateral
joint research projects.

\appendix

\section{Instanton corrections at integer $k$}\label{app:inst}
In this Appendix, we summarize the instanton corrections at some integer $k$.
\subsection{Fundamental representation}\label{app:fund}
For the fundamental representation with $k=3,4,6,8,12$ we find
\begin{align}
\begin{aligned}
 e^{-\frac{2\mu}{3}}\cW_1(k=3)&=
\left(\frac{2\mu}{3\rt{3}\pi}+\frac{2}{9}\right)+\frac{4 \mu }{3 \sqrt{3} \pi }e^{-\frac{4\mu}{3}}
+\left(-\frac{2 \mu }{3 \sqrt{3} \pi }-\frac{14}{9}\right)e^{-\frac{8\mu}{3}}
+\left(\frac{-96\mu+18}{27\rt{3}\pi}+\frac{40}{27}\right)e^{-4\mu} \\
&+\left(\frac{-146\mu+12}{9\rt{3}\pi}-\frac{70}{9}\right)e^{-\frac{16\mu}{3}}
+\left(\frac{48\mu-2}{3\rt{3}\pi}+\frac{320}{9}\right)e^{-\frac{20\mu}{3}}
+\left(\frac{652\mu-177}{27\rt{3}\pi}-\frac{4892}{81}\right)e^{-8\mu},\\
e^{-\frac{\mu}{3}}\cW_1(k=6)&=
\left(\frac{\mu}{3\rt{3}\pi}-\frac{1}{9}\right)+\frac{2\mu}{3\rt{3}\pi}e^{-\frac{2\mu}{3}}
+\left(-\frac{ \mu }{3 \sqrt{3} \pi }+\frac{7}{9}\right)e^{-\frac{4\mu}{3}}
+\left(\frac{-48\mu+18}{27\rt{3}\pi}-\frac{20}{27}\right)e^{-2\mu}\\
&+\left(\frac{-73\mu+12}{9\rt{3}\pi}+\frac{35}{9}\right)e^{-\frac{8\mu}{3}}
+\left(\frac{24\mu-2}{3\rt{3}\pi}-\frac{160}{9}\right)e^{-\frac{10\mu}{3}}
+\left(\frac{326\mu-177}{27\rt{3}\pi}+\frac{2446}{81}\right)e^{-4\mu},\\
e^{-\frac{\mu}{2}}\cW_1(k=4)&=\frac{\mu}{4\pi}
+\frac{\mu}{2\pi}e^{-\mu}+\frac{3\mu-1}{2\pi}e^{-2\mu}+\frac{5\mu-1}{\pi}e^{-3\mu}
+\frac{74\mu-21}{4\pi}e^{-4\mu}\\
& +\frac{130\mu-29}{2\pi}e^{-5\mu}+\frac{1518\mu-403}{6\pi}e^{-6\mu},\\
 e^{-\frac{\mu}{4}}\cW_1(k=8)&=
 \left(\frac{\mu}{4\rt{2}\pi}-\frac{1}{4\rt{2}}\right)
 +\frac{\mu}{2\rt{2}\pi}e^{-\frac{\mu}{2}}
 +\left(-\frac{\mu}{4\rt{2}\pi}+\frac{5}{4\rt{2}}\right)e^{-\mu}\\
 &+\left(\frac{\mu}{2\rt{2}\pi}-2\rt{2}\right)e^{-\frac{3\mu}{2}}
+\left(\frac{\mu -4}{8 \sqrt{2} \pi }+\frac{79}{8 \sqrt{2}}\right)e^{-2\mu}\\
&+\left(\frac{41 \mu-4}{4 \sqrt{2} \pi }-16\rt{2}\right)e^{-\frac{5\mu}{2}}
+\left(\frac{-139 \mu+4}{8 \sqrt{2} \pi }+\frac{791}{8\rt{2}}\right)e^{-3\mu},\\
e^{-\frac{\mu}{6}}\cW_1(k=12)&=\frac{\mu}{6\pi}
\Big(1+2e^{-\frac{\mu}{3}}-e^{-\frac{2\mu}{3}}+2e^{-\mu}-7e^{-\frac{4\mu}{3}}+30e^{-\frac{5\mu}{3}}\Big)\\
 &-\frac{1}{2\rt{3}}\Big(1-\frac{13}{3}e^{-\frac{2\mu}{3}}+\frac{46}{3}e^{-\mu}-\frac{185}{3}e^{-\frac{4\mu}{3}}
 +\frac{730}{3}e^{-\frac{5\mu}{3}}\Big).
\end{aligned}
\end{align}

By rewriting the above expansions in terms of $\mu_\text{eff}$, we find
\begin{align}
 \begin{aligned}
 e^{-\frac{2\mu_\text{eff}}{3}}
  \cW_1(k=3)&=
\frac{2\mu_\text{eff}}{3\rt{3}\pi}\Big(1 + 2 Q_w - Q_w^2 - 6 Q_w^3 - 23 Q_w^4 + 22 Q_w^5 + 19 Q_w^6\Big)\\
&\qquad+\frac{2}{9}\Big(1 - 7 Q_w^2 + 6 Q_w^3 - 35 Q_w^4 + 146 Q_w^5 - 249 Q_w^6\Big),\\
e^{-\frac{\mu_\text{eff}}{3}}
  \cW_1(k=6)&=\frac{\mu_\text{eff}}{3\rt{3}\pi}\Big(1 + 2 Q_w - Q_w^2 - 6 Q_w^3 - 23 Q_w^4 + 22 Q_w^5+19Q_w^6\Big)\\
  &\qquad-\frac{1}{9}\Big(1 - 7 Q_w^2 + 6 Q_w^3 - 35 Q_w^4 + 146 Q_w^5-249 Q_w^6\Big),\\
  e^{-\frac{\mu_\text{eff}}{6}}
  \cW_1(k=12)&=\frac{\mu_\text{eff}}{6\pi}\Big(1 + 2 Q_w - Q_w^2 + 2 Q_w^3 - 7 Q_w^4 + 30 Q_w^5 - 117 Q_w^6 + 476 Q_w^7\Big)\\
  &-\frac{1}{2\rt{3}}\left(1 - \frac{13}{3} Q_w^2 +
  \frac{46}{3} Q_w^3 - \frac{185}{3} Q_w^4+ \frac{730}{3} Q_w^5 -
 \frac{2651}{3} Q_w^6 + 3146 Q_w^7\right),\\
e^{-\frac{\mu_\text{eff}}{4}}
  \cW_1(k=8)&=\frac{\mu_\text{eff}}{4\rt{2}\pi}
\Big(1 + 2 Q_w - Q_w^2 + 2 Q_w^3 + Q_w^4 + 40 Q_w^5 - 68 Q_w^6\Big)\\
&\quad-\frac{1}{4\rt{2}}
\Big(1 - 5 Q_w^2 + 16 Q_w^3 - 39 Q_w^4 + 128 Q_w^5 - 388 Q_w^6\Big),\\
  e^{-\frac{\mu_\text{eff}}{2}}
  \cW_1(k=4)&=
  \frac{\mu_\text{eff}}{4\pi}\Big(1+2 Q_w + 7  Q_w^2 + 18  Q_w^3 + 57 Q_w^4  + 160  Q_w^5+516Q_w^6\Big).
 \end{aligned}
\end{align}

\subsection{Winding number $n=2$}
For the winding number $n=2$ with $k=6,8,12$ we find
\begin{align}
 \begin{aligned}
  e^{-\frac{4\mu_\text{eff}}{k}}\cW_2(k=6)
  &=\frac{\mu_\text{eff}}{3\rt{3}\pi}(1 - Q_w - 12 Q_w^3 + 18 Q_w^4 - 20 Q_w^5 + 156 Q_w^6 - 292 Q_w^7)\\
  & +\frac{2}{9}(Q_w - Q_w^2 + 6 Q_w^3 - 23 Q_w^4 + 40 Q_w^5 - 133 Q_w^6 + 408 Q_w^7),\\
  e^{-\frac{4\mu_\text{eff}}{k}}\cW_2(k=8)&=
  \frac{\mu_\text{eff}}{8\pi}(1 - 2 Q_w + 2 Q_w^2 - 8 Q_w^3 + 35 Q_w^4 - 88 Q_w^5)\\
  &-\frac{1}{8}(1 - 2 Q_w + 2 Q_w^2 - 8 Q_w^3 + 39 Q_w^4 - 96 Q_w^5),\\
   e^{-\frac{4\mu_\text{eff}}{k}}\cW_2(k=12)&=
  \frac{\mu_\text{eff}}{6\rt{3}\pi}(1 - 3 Q_w + 4 Q_w^2 - 12 Q_w^3+50Q_w^4-204Q_w^5 )\\
  &-\frac{2}{9}\left(1 - \frac{3}{2} Q_w +  \frac{3}{2}Q_w^2 - 6 Q_w^3+  \frac{55}{2}Q_w^4 -120Q_w^5\right).
 \end{aligned}
\label{eq:w2-int-k}
\end{align}

\subsection{Winding number $n=3$}
For the winding number $n=3$ with $k=8,12$ we find
\begin{align}
 \begin{aligned}
  e^{-\frac{6\mu_\text{eff}}{k}}\cW_3(k=8)&=
  \frac{\mu_\text{eff}}{4\rt{2}\pi}(1 - Q_w + 3 Q_w^2 - 7 Q_w^3 + 32 Q_w^4 - 73 Q_w^5)\\
  &+\frac{1}{4\rt{2}}(Q_w - 4 Q_w^2 + 11 Q_w^3 - 36 Q_w^4 + 105 Q_w^5),\\
  e^{-\frac{6\mu_\text{eff}}{k}}\cW_3(k=12)&=\frac{\mu_\text{eff}}{12\pi}(1-4Q_w+12Q_w^2-34Q_w^3+120Q_w^4-460Q_w^5)\\
  &-\frac{1}{3\rt{3}}(1-2Q_w+6Q_w^2-20Q_w^3+80Q_w^4-310Q_w^5).
 \end{aligned}
\end{align}

\subsection{Winding number $n=4$}
For the winding number $n=4$ with $k=12$ we find
\begin{align}
 \begin{aligned}
   e^{-\frac{8\mu_\text{eff}}{k}}\cW_4(k=12)&=
  \frac{\mu_\text{eff}}{6\rt{3}\pi}(1 - 3 Q_w + 10 Q_w^2 - 36 Q_w^3+129Q_w^4)\\
  &-\frac{1}{6}\left(1 - 2 Q_w +  \frac{23}{3}Q_w^2 - 30 Q_w^3+  \frac{335}{3}Q_w^4\right).
 \end{aligned}
\end{align}

\subsection{Imaginary part of $W_{(1,1)}$}\label{app:Im}
For the imaginary part of $W_{(1,1)}$ with $k=6,8,12$ we find
\begin{align}
 \begin{aligned}
  e^{-\frac{2\mu_\text{eff}}{3}}W_{(1,1)}^{\text{Im}}(k=6)&=\frac{\mu_\text{eff}}{18\pi}(1+3Q_w-8Q_w^2+4Q_w^3-52Q_w^4+140Q_w^5)\\
  &+\frac{1}{9\rt{3}}(1+Q_w-Q_w^2-2Q_w^3-25Q_w^4+52Q_w^5),\\
  e^{-\frac{\mu_\text{eff}}{2}}W_{(1,1)}^{\text{Im}}(k=8)&=\frac{\mu_\text{eff}}{16\pi}(1+2Q_w-6Q_w^2+8Q_w^3-13Q_w^4+88Q_w^5)\\
  &+\frac{1}{8}(0+Q_w+0\cdot Q_w^2-4Q_w^3+0\cdot Q_w^4+16Q_w^5),\\
e^{-\frac{\mu_\text{eff}}{3}}W_{(1,1)}^{\text{Im}}(k=12)&=\frac{\mu_\text{eff}}{12\pi}(1+Q_w-4Q_w^2+4Q_w^3-6Q_w^4+20Q_w^5)\\
&-\frac{1}{6\rt{3}}(1-3Q_w-Q_w^2+16Q_w^3-55Q_w^4+172Q_w^5).
 \end{aligned}
\end{align}

\subsection{Real part of $W_{(1,1)}$}
For the real part of $W_{(1,1)}$ with $k=6,8$ we find
\begin{align}
 \begin{aligned}
   e^{-\frac{2\mu_{\text{eff}}}{3}}W_{(1,1)}^{\text{Re}}(k=6)&=
-\frac{\mu_\text{eff}^2}{54\pi^2}(1 + 7 Q_w - 4 Q_w^2 + 32 Q_w^3-222Q_w^4+300Q_w^5-1196Q_w^6)\\
&-\frac{2\mu_\text{eff}}{27\rt{3}\pi}(1 - Q_w + 7 Q_w^2 + 22 Q_w^3-9Q_w^4-160Q_w^5-681Q_w^6)\\
&+\frac{1}{648}\Bigl(11 - 75 Q_w + 284 Q_w^2 - 696 Q_w^3+1302Q_w^4-6076Q_w^5+12236Q_w^6\Bigr)\\
&-\frac{1}{6\rt{3}\pi}\Bigl(1 - Q_w - 16Q_w^3+22Q_w^4-20Q_w^5+224Q_w^6\Bigr),\\
 e^{-\frac{\mu_{\text{eff}}}{2}}W_{(1,1)}^{\text{Re}}(k=8)&=
-\frac{\mu_\text{eff}^2}{64\pi^2}(1 + 6 Q_w - 2 Q_w^2 + 32 Q_w^3 - 89 Q_w^4+368Q_w^5)\\
&+\frac{\mu_\text{eff}}{16\pi}(Q_w - 4 Q_w^2 + 8 Q_w^3 - 8 Q_w^4+44Q_w^5)\\
&+\frac{1}{16}(1 - 3 Q_w + 11 Q_w^2 - 38 Q_w^3 + 105 Q_w^4-312Q_w^5)\\
&-\frac{1}{16\pi}(1 - 2 Q_w + 2 Q_w^2 - 8 Q_w^3 + 39 Q_w^4-96Q_w^5).
 \end{aligned}
\label{eq:Wre-k68}
\end{align}


\end{document}